\documentclass[preprint,3p,times,authoryear]{elsarticle}

\usepackage{epsfig}
\usepackage{multirow}
\usepackage{multicol}
\usepackage{listings}
\usepackage{comment}
\usepackage{amssymb}
\usepackage{amsmath}
\usepackage{hyperref}
\usepackage{color, xcolor, colortbl}

\definecolor{mygreen}{rgb}{0,0.6,0}
\definecolor{mygray}{rgb}{0.47,0.47,0.33}
\definecolor{myorange}{rgb}{0.8,0.4,0}
\definecolor{mywhite}{rgb}{0.98,0.98,0.98}
\definecolor{myblue}{rgb}{0.01,0.61,0.98}
\definecolor{backcolour}{rgb}{0.96,0.96,0.94}
\definecolor{yellowgreen}{rgb}{0.57,0.82,0.31}
\definecolor{codegreen}{HTML}{3D7A7A}
\definecolor{codered}{HTML}{B00040}
\definecolor{codeblue}{HTML}{0000FF}
\definecolor{richelectricblue}{rgb}{0.03, 0.57, 0.82}

\lstdefinestyle{yaml}{
     basicstyle=\color{blue}\footnotesize,
     rulecolor=\color{black},
     string=[s]{'}{'},
     stringstyle=\color{blue},
     comment=[l]{:},
     commentstyle=\color{black},
     morecomment=[l]{-},
     frame=single,
 }
\usepackage[ruled,vlined,linesnumbered]{algorithm2e}
\DontPrintSemicolon

\AtBeginDocument{%
  \providecommand\BibTeX{{%
    \normalfont B\kern-0.5em{\scshape i\kern-0.25em b}\kern-0.8em\TeX}}}

\graphicspath{{./images/}} 

\journal{Elsevier}

\begin{document}
\begin{frontmatter}

\title{CONTINUUM: Detecting APT Attacks through Spatial-Temporal Graph Neural Networks}

\author[LCSI]{Atmane Ayoub Mansour Bahar\corref{corr}
}
\ead{ja\_mansourbahar@esi.dz}

\affiliation[LCSI]{organization={LCSI, École Nationale Supérieure d'Informatique (ESI ex. INI)},
             city={Oued Smar},
             postcode={16058},
             state={Algiers},
             country={Algeria}}

\author[LCSI]{Kamel Soaïd Ferrahi\corref{corr}
}
\ead{jk_ferrahi@esi.dz}

\author[ERIC]{Mohamed-Lamine Messaï}
\ead{mohamed-lamine.messai@univ-lyon2.fr}

\author[LIRIS]{Hamida Seba}
\ead{{hamida.seba@univ-lyon1.fr}}

\author[LCSI]{Karima Amrouche}
\ead{k_amrouche@esi.dz}

\affiliation[ERIC]{organization={ERIC, Université Lyon 2},
             city={Bron},
             state={Lyon},
             country={France}}

\affiliation[LIRIS]{organization={Universite Claude Bernard Lyon 1, CNRS, INSA Lyon, LIRIS, UMR5205},
             city={69622 Villeurbanne},
             country={France}}

\cortext[corr]{Both authors contributed equally to this research.}
\begin{abstract}
Advanced Persistent Threats (APTs) represent a significant challenge in cybersecurity due to their sophisticated and stealthy nature. Traditional Intrusion Detection Systems (IDS) often fall short in detecting these multi-stage attacks. Recently, Graph Neural Networks (GNNs) have been employed to enhance IDS capabilities by analyzing the complex relationships within networked data. However, existing GNN-based solutions are hampered by high false positive rates and substantial resource consumption.
In this paper, we present a novel IDS designed to detect APTs using a Spatio-Temporal Graph Neural Network Autoencoder. Our approach leverages spatial information to understand the interactions between entities within a graph and temporal information to capture the evolution of the graph over time. This dual perspective is crucial for identifying the sequential stages of APTs. Furthermore, to address privacy and scalability concerns, we deploy our architecture in a federated learning environment. This setup ensures that local data remains on-premise while encrypted model-weights are shared and aggregated using homomorphic encryption, maintaining data privacy and security.
Our evaluation shows that this system effectively detects APTs with lower false positive rates and optimized resource usage compared to existing methods, highlighting the potential of spatio-temporal analysis and federated learning in enhancing cybersecurity defenses. 
\end{abstract}

\begin{keyword}
Graph Neural Networks \sep Intrusion Detection \sep Advanced Persistent Threats \sep Graph Attention Networks \sep Gated Recurrent Unit \sep Spatial-Temporal Analysis \sep Federated Learning \sep Homomorphic Encryption
\end{keyword}

\end{frontmatter}

\section{Introduction}
\label{sec:intro}
In recent years, the frequency and sophistication of cyber-attacks have escalated, posing significant challenges to the cybersecurity community. Among these threats, Advanced Persistent Threats (APTs) stand out due to their stealthy, prolonged, and multi-stage nature. APTs are highly targeted attacks typically orchestrated by well-funded adversaries, aiming to gain and maintain unauthorized access to a network while evading detection for extended periods. These multi-stage attacks often involve a series of coordinated steps, including initial intrusion, lateral movement, privilege escalation, and data exfiltration.

Traditional Intrusion Detection Systems (IDS) have been pivotal in safeguarding network security by monitoring and analyzing network traffic for signs of malicious activities. However, the sophisticated tactics employed by APTs often render these traditional systems inadequate. APTs can adapt and evolve their techniques to bypass conventional security measures, involving more advanced detection mechanisms.

One promising approach to understanding and detecting these complex threats is through the analysis of provenance graphs. Provenance graphs capture the history and lineage of data and events within a system, detailing the interactions and dependencies between various system-entities. These graphs provide a comprehensive view of the sequence and causality of events, making them invaluable for identifying suspicious-activities and potential security breaches. 

Graph Neural Networks (GNNs) have gained prominence in recent years for their ability to process and analyze graph-structured data. GNNs excel in capturing deep information and correlations within graphs, making them particularly suited for understanding the behavior of system entities as represented in provenance graphs. Using GNNs, it is possible to uncover intricate patterns and relationships that traditional methods might miss, thus enhancing the detection of APTs.


However, traditional IDS and even recent GNN-based solutions often fall short in effectively detecting APTs because they do not capture the time dimension in these attacks. 
Thus, in this paper, we propose a novel IDS based on a Spatial-Temporal Graph-Autoencoder. Our approach leverages spatial information to understand the interactions between entities within a graph and temporal information to capture the evolution of the graph over time. It starts by processing provenance graphs from public benchmark datasets, using a method to generate and compress snapshots, making the datasets time dependent. We leverage then a Graph Attention Network (GAT)-based Autoencoder to extract spatial information from these provenance graphs, enabling the understanding of interactions between entities. 
Additionally, we employ Gated Recurrent Unit (GRU) gates to store and track the evolution of the provenance graph over time, crucial for detecting the multi-stage nature of APTs executed in different stages separated by time intervals. This dual perspective of spatial and temporal information significantly improves the accuracy and effectiveness of APT detection.

The remainder of the paper is organized as follows: Section \ref{sec:defs} provides definitions of key concepts and presents the threat model. In Section \ref{sec:motiv}, we present the motivations underlying this work.  
 Section \ref{sec:related-work} reviews related work. Then, Section \ref{sec:our-approach} details our system design. 
 In Section \ref{sec:Perf}, we present the results of the evaluation of our findings. 
 Finally, Section \ref{sec:conclusion} concludes the paper and outlines some future research directions.

\section{Preliminaries}
\label{sec:defs}
In this section, we define 
the key concepts used in this work as well as the threat model. 
\subsection{Definitions}
The key concepts defined are as follows: 
Advanced Persistent Threats, provenance graphs, and a specific category of Graph Neural Networks, called Spatial Convolutional Graph Neural Networks.

\noindent \textbf{Advanced Persistent Threats (APts): } APTs are sophisticated and prolonged cyberattacks \citep{alshamrani2019survey} aimed at stealing sensitive information, disrupting operations, or causing damage to targeted entities \citep{techtargetWhatAdvanced}. The primary characteristics of APTs include their advanced nature, which involves complex techniques, tools, and exploits to compromise the target, their persistence, which ensures that they remain within the target network for an extended period to achieve long-term objectives, and the significant threat they pose, often carried out by highly skilled adversaries with substantial resources, such as state-sponsored or well-funded organizations.

Unlike typical cyberattacks, APTs involve multiple stages \citep{quintero2020new, ussath2016advanced, sexton2015attack, vukalovic2015advanced, Swisscom2019} and employ various techniques to maintain access and evade detection over long periods, each stage involving specific actions and objectives. For example, some common APT stages are as follows:
\begin{itemize}
    \item \textbf{Initial Intrusion:} Exploiting vulnerabilities or using social engineering to gain access to the target network.
    \item \textbf{Establishing Foothold:} Deploying malware or backdoors to maintain access.
    \item \textbf{Lateral Movement:} Moving laterally within the network to identify high-value targets and gain additional privileges.
    \item \textbf{Privilege Escalation:} Obtaining higher-level access to perform more significant actions.
    \item \textbf{Data Exfiltration or Attack Execution:} Stealing sensitive information or executing destructive actions.
    \item \textbf{Persistence:} Ensuring long-term access and evading detection through various means.
\end{itemize}

\begin{figure}[h]
\centering
\includegraphics[width=0.7\textwidth]{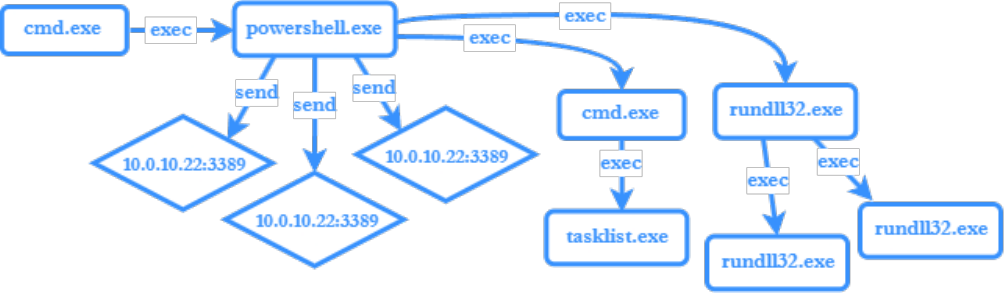}
\caption{Provenance graph of an APT attack \citep{Hassan2020TacticalPA}}
\label{fig:provenance}
\end{figure}

\noindent \textbf{Provenance Graphs: } A provenance graph is a directed acyclic graph $G=(V,E)$ that represents the history of data objects, where $V$ is the set of nodes representing system entities (files, users, tasks, processes, ect..), and $E$ is the set of directed edges representing the interactions among these entities. An example of an APT attack's provenance graph is illustrated in Figure \ref{fig:provenance}.

Provenance graphs are valuable for intrusion detection because they capture detailed activity, providing a comprehensive record of system activities that makes it easier to identify anomalous behavior \citep{acar2010graph}. They also reveal hidden dependencies, highlighting relationships between different entities, which can be crucial for detecting lateral-movements and other APT stages. In addition, they facilitate forensic analysis by allowing investigators to trace back the origin and propagation of an attack.

\noindent \textbf{Graph Neural Networks (GNNs): }
GNNs  are a class of neural networks specifically designed to operate on graph-structured data \citep{liu2022introduction}. A graph \( G = (V, E) \) consists of a set of nodes \( V \) and edges \( E \), where each edge \( (i, j) \in E \) represents a relationship between nodes \( i \) and \( j \). Each node \( v_i \in V \) may also have a feature vector \( x_i \in \mathbb{R}^d \), representing the attributes of the node. GNNs aim to learn node representations by iteratively aggregating information from their neighbors. At each layer \( l \) of the GNN, the representation of node \( v_i \) is updated by aggregating the feature vectors of its neighboring nodes \( \mathcal{N}(v_i) \), followed by a non-linear transformation.

The general formulation of a GNN layer can be written as:
\begin{equation}
h_i^{(l+1)} = \sigma \left( W^{(l)} h_i^{(l)} + \sum_{j \in \mathcal{N}(i)} W_{\text{neigh}}^{(l)} h_j^{(l)} \right)
\end{equation}
where 
\( h_i^{(l)} \) is the feature vector of node \( i \) at layer \( l \),
 \( \mathcal{N}(i) \) represents the set of neighbors of node \( i \),
 \( W^{(l)} \) and \( W_{\text{neigh}}^{(l)} \) are learnable weight-matrices for node \( i \) and its neighbors, respectively, and 
 \( \sigma \) is a non-linear activation function (e.g., ReLU).
This equation describes how each node's representation is updated by combining its current representation \( h_i^{(l)} \) with the aggregated information from its neighbors \( h_j^{(l)} \). The learned node embeddings \( h_i^{(L)} \) after \( L \) layers can then be used for various downstream tasks such as node classification, link prediction, and graph classification.

\cite{wu2020comprehensive} categorized GNNs based on their architecture and message-passing mechanisms into four primary types of GNNs: Recurrent GNNs (recGNN) \citep{li2015gated, gallicchio2010graph, scarselli2008graph, dai2018learning}, Convolutional GNNs (convGNN) \citep{velivckovic2017graph, hamilton2017inductive, xu2018powerful, gilmer2017neural, velickovic2019deep, defferrard2016convolutional, kipf2016semi, li2018adaptive, zhuang2018dual, levie2018cayleynets}, Graph Autoencoders (GAEs) \citep{kipf2016variational, wang2016structural, hajiramezanali2019variational, bojchevski2018netgan, tu2018deep}, and Spatial-Temporal GNNs (STGNN) \citep{chen2023st, seo2018structured, wang2022attention, yan2018spatial, li2023stgate, xu2020spatial}. Each category of GNN serves a different purpose, offering unique advantages depending on the task at hand.
\begin{itemize}
\item Recurrent Graph Neural Networks (recGNNs): 
recGNNs incorporate recurrent connections within their architecture to iteratively update node representations over multiple steps. The hidden state of each node is updated based on its previous hidden states and those of its neighboring nodes using a recurrent neural network (RNN) layer such as GRU or LSTM. This iterative process allows recGNNs to capture sequential information exchange between nodes in a graph.

A general update function for recGNNs is expressed as:
\[
h^{(t)}(v) = f_{\text{RNN}} \left( W \cdot h^{(t-1)}(v), \sum_{u \in \mathcal{N}(v)} h^{(t-1)}(u) \right)
\]
where 
\( h^{(t)}(v) \) is the hidden state of node \( v \) at time step \( t \),
\( f_{\text{RNN}} \) is the recurrent neural network function,
\( \mathcal{N}(v) \) represents the neighboring nodes of node \( v \), and 
\( W \) is the learnable weight matrix.



    \item Convolutional Graph Neural Networks (ConvGNNs): 
ConvGNNs extend the concept of convolutional operations from traditional grid-structured data (such as images) to irregular graph-structured data. They can be categorized into two main types: Spatial Convolutional GNNs (SpGNNs) \citep{velivckovic2017graph, hamilton2017inductive, xu2018powerful, gilmer2017neural, velickovic2019deep} and Spectral Convolutional GNNs (ScGNNs) \citep{defferrard2016convolutional, kipf2016semi, li2018adaptive, zhuang2018dual, levie2018cayleynets}.
\begin{itemize}

\item Spatial Convolutional GNNs:
Spatial convolutional GNNs perform convolutions by aggregating information from a node’s neighbors \citep{mbernsteGraphConvolutional}. The key idea is to iteratively update the representation of a node by combining its features with those of its neighbors. This process captures the local structure and features of the graph.

Formally, the update rule for the node representations in spatial convolutional GNNs can be described as follows:
\begin{equation}
h_i^{(k)} = \sigma \left( W^{(k)} h_i^{(k-1)} + \sum_{j \in \mathcal{N}(i)} W_{\text{neigh}}^{(k)} h_j^{(k-1)} \right)
\end{equation}

where 
 \( h_i^{(k)} \) is the feature vector of node \( i \) at the \( k \)-th layer,
  \( W^{(k)} \) and \( W_{\text{neigh}}^{(k)} \) are learnable weight matrices for the node itself and its neighbors, respectively, 
  \( \mathcal{N}(i) \) denotes the set of neighbors of node \( i \), and 
 \( \sigma \) is an activation function such as ReLU.

The above equation signifies that the new representation of a node $hi(k)$ is obtained by combining its previous representation $hi(k-1)$ with the representations of its neighbors $hj(k-1)$, weighted by the matrices $W(k)$ and $Wneigh(k)$. This aggregation step effectively captures the spatial dependencies and relationships within the graph.

Some well-known examples of SpGNNs are Graph Attention Networks \citep{velivckovic2017graph}, GraphSAGE \citep{hamilton2017inductive}, and Graph Isomorphism Networks \citep{xu2018powerful}. By iteratively applying this update rule across multiple layers, these GNNs can capture higher-order neighborhood information, enabling them to learn rich representations of the nodes and their interactions. This makes them particularly well-suited for tasks such as node classification, link prediction, and graph classification.

\item Spectral Convolutional GNNs: Spectral Convolutional GNNs perform convolutions in the spectral domain by leveraging the eigenvalues and eigenvectors of the graph Laplacian \citep{tkipfPowerfulGraph}. The central idea is to transform node features into the spectral domain, apply spectral filters, and then map them back to the spatial domain. This process captures global graph structure while maintaining efficiency using approximations like Chebyshev polynomials \citep{mason2002chebyshev}.

Formally, the spectral convolution can be expressed as: \begin{equation} h_v^{(l+1)} = U \cdot \text{diag}(\theta^{(l)}) \cdot U^\top \cdot h_v^{(l)} \end{equation}

where 
\( h_i^{(k)} \) and \( h_i^{(k+1)} \) are the feature vectors of node \( i \) at the \( k \)-th and \( k+1 \)-th layers respectively, 
\( U \) is the matrix of eigenvectors of the graph Laplacian,
\( \theta^{(k)} \) represents the spectral filter parameters for layer \( k \),
\( diag(\theta^{(k)}) \) is a diagonal matrix formed from layer \( \theta^{(k)} \).

The equation indicates that the updated node features are obtained by filtering the node representations in the spectral domain and transforming them back to the original domain. This process allows ScGNNs to effectively capture global graph properties and relationships.

Examples of ScGNNs include ChebNet \citep{defferrard2016convolutional}, which uses Chebyshev polynomials to approximate spectral filters, and Graph Convolutional Networks (GCN) \citep{kipf2016semi}, which simplify spectral convolution by truncating the spectral decomposition. These models excel in learning representations for tasks such as node classification, link prediction, and graph classification, particularly in structured domains.
\end{itemize}

\item Graph Autoencoders (GAEs):
GAEs are a class of GNNs that use an autoencoder framework to encode nodes into embeddings and learn low-dimensional representations of graphs. These models are designed to capture the essential structural properties of a graph while allowing for compression and reconstruction of the graph structure. GAEs define an unsupervised reconstruction error between the original and predicted adjacency matrices, ensuring that the model learns meaningful representations \citep{ward2022practical}.

The general encoding process can be expressed as:
\begin{equation}
z(v) = f \left( \sum_{u \in \mathcal{N}(v)} \frac{1}{\sqrt{d(u) d(v)}} W \cdot x(u) \right)
\end{equation}
where 
    \( z(v) \) is the embedding of node \( v \),
    \( d(v) \) and \( d(u) \) are the degrees of nodes \( v \) and \( u \), 
    \( W \) is the weight matrix, and \( x(u) \) is the input feature vector.
Examples of GAEs include Variational Graph Autoencoders (VGAE) \citep{kipf2016variational} and Structure Deep Network Embeddings (SDNE) \citep{wang2016structural}.

\item Spatial-Temporal Graph Neural Networks (STGNNs): 
Spatial-Temporal Graph Neural Networks (STGNNs) extend GNNs to handle temporal dynamics by integrating spatial and temporal information. These models are particularly useful for spatio-temporal data, such as traffic forecasting or time-evolving networks. STGNNs use spatial layers to aggregate information from neighboring nodes within the same time step and temporal layers to capture dependencies across time steps.

A generalized update equation for STGNNs is:
\begin{equation}
h^{(t)}(v) = f \left( h^{(t-1)}(v), 
 \text{AGG}_{\text{spatial}}^{(t)} \left( h^{(t-1)}(u) : u \in \mathcal{N}(v) \right), \text{AGG}_{\text{temporal}}^{(t)} \left( h^{(t)}(v) \right) \right)
\end{equation}
where 
 \( \text{AGG}_{\text{spatial}}\) is the Spatial aggregation-function, and 
\( \text{AGG}_{\text{temporal}}\) is the Temporal aggregation-function.

Examples include Gated Spatio-Temporal Graph Convolutional Networks (ST-GCN) \citep{yan2018spatial} and Temporal Graph Networks (TGN) \citep{rossi2020temporal}.
\end{itemize}
\subsection{Threat Model}
In the proposed model, we assume that the adversary has advanced capabilities, including the ability to breach the system perimeter through various attack vectors such as phishing (social engineering) or exploitation of unpatched vulnerabilities. Once inside the system, the attacker seeks to maintain persistent access by installing back-doors or leveraging malware, moving laterally within the network to escalate privileges, and carrying out the attack over an extended period without being detected.
We also assume that the attacker can compromise both hosts and servers within the organization. However, he leaves behind him abnormal system events and entity interactions, which are represented in the provenance graph, and analyzed to detect anomalies indicative of an APT attack.

\section{Motivation}
\label{sec:motiv}
The research presented in this paper is driven by several key-motivations regarding the nature of Advances Persistent Threats and their detection by exploiting GNNs. In this section, we present an overview of these motivations: 
\begin{itemize}
\item \textbf{Addressing the stealthiness od APTs:} 
Advanced Persistent Threats are highly stealthy and sophisticated, often blending their activities with legitimate network traffic to avoid detection. Traditional anomaly detection methods struggle with the vast amount of data and the subtlety of these anomalies \citep{ADNAN2023Forensic, Zimba2020Modeling, Lu2019A, Su2017APT}. To overcome this challenge, we need a solution that allows learning compressed representations of data, and effectively distinguishing normal behavior from anomalies. For this purpose, we propose to utilize graph-autoencoders. Autoencoders are powerful tools for learning compressed representations of data \citep{Pawar2019Assessment, Bianchi2019Learning, Shen2016Lossless}. By training an autoencoder on benign system behavior, the model learns to generate precise embeddings for nodes and graphs. When the autoencoder encounters anomalous data, such as APT activity, it produces higher reconstruction errors, thereby facilitating detection of attacks. 
This approach not only improves detection accuracy but also enhances the efficiency of the IDS by reducing the dimensionality of the data.

\item \textbf{Encompassing the relationships between system entities: } 
APTs 
often exploit the relationships between system entities (users, files, processes, ... etc.) to hide their malicious activities within benign system operations and propagate within the system. To effectively detect these threats, it is essential to deeply understand the relationships represented in provenance graphs \citep{yan2022deepro, Lv2022A}, and extract the correlations between their nodes and edges. For that, we need models that focus on the information transmitted between a node and its neighbourhood, called the spatial information, in order to capture the relationships and interactions between entities, which can reveal hidden patterns indicative of APT activities. 

The solution that we propose in this paper is implementing a Spatial convolutional GNN, such as Graph Isomorphism Networks (GIN) \citep{xu2018powerful}, Graph Attention Networks (GAT) \citep{velivckovic2017graph}, and GraphSAGE \citep{hamilton2017inductive}. This type of GNNs is particularly suited for modelling the spatial dynamics of the network (messages passed in a nodes' neighbourhood), and
understanding intricate correlation between system entities \citep{sahili2023spatio, wu2020comprehensive}.

\item \textbf{Capturing time span of APTs: }
APTs tend to be prolonged and encompass different stages, including initial intrusion, lateral movement, privilege escalation, and data exfiltration. These stages can occur over extended periods, making it crucial to identify the time-dependencies of APT campaigns \citep{cheng2023kairos, Zhu2021General, Ghafir2019Hidden}. To detect these temporal patterns, we integrate machine-learning models that possess memory capabilities and can retain information separated by time. Models such as Recurrent Neural Networks (RNNs) \citep{rumelhart1986learning}, Long Short-Term Memory networks (LSTMs) \citep{hochreiter1997long}, and Gated Recurrent Units (GRUs) \citep{cho2014learning}, are effective in storing and processing sequential-data, allowing our system to track the evolution of APTs across different stages.

\item \textbf{Encompassing the dynamic aspects of APTs: }
APT attackers usually aim to compromise a whole system rather than a specific host, making it crucial to have a model that understands the behavior of the organization as a whole \citep{mansour2024fedhe}. Local models often struggle to generalize to new, dynamic, and evolving APT attacks because they are trained on data from a single host, limiting their ability to detect threats that target an entire organization \citep{Zimba2020Modeling, Li2023Few-shot}. 
Federated Learning (FL), for instance, offers a solution to these issues by enabling collaborative training across different hosts, without having to share their raw data \citep{son2023xfedgraph}. This decentralized approach not only enhances the detection of organization-wide threats but also reduces training time by leveraging the computational resources of multiple clients \citep{mansour2024fedhe, son2023xfedgraph}. By collaboratively training on diverse data from various hosts, the model can generalize better to new APTs and provide a more comprehensive defense mechanism.

Federated solutions can also be deployed in a large-scale environment, where different companies or state-related organisations collaborate with their respective model weights to build an Intrusion Detection System that can generalise to various system-behaviours, and effectively detect new APT attacks. This architecture plays the role of a defensive alliance among organisations against targeted and sophisticated attacks.

However, Federated Learning introduces new privacy concerns, as model weights are shared between clients and servers, making them vulnerable to data leakage \citep{shanmugarasa2023systematic}.

\item \textbf{Addressing Privacy Concerns: }
While Federated Learning enhances collaborative training, it also raises significant privacy issues. Transmitting model weights in clear text can expose them to Man-in-the-Middle (MitM) attacks, where an adversary can intercept and reverse-engineer the data of clients. In addition, an attacker could target the central server to obtain model weights in what is known as Inference Attacks \citep{krumm2007inference}, compromising the privacy of all participating clients. To mitigate these risks, a solution that can be employed is Homomorphic Encryption \citep{yi2014homomorphic, doan2023survey}, which allows computations to be performed on encrypted data without the need to decrypt it. This ensures that even if model weights are intercepted, the data remains secure, and the server cannot access the raw weights. It also opens perspectives on implementing Intrusion Detection Systems with untrusted-servers, or serveless architectures \citep{Zhong2022MPC-Based, Baughman2022Exploring, Savazzi2019Federated}.

\end{itemize}
\noindent Our research combines these advanced techniques to develop a robust Intrusion Detection System capable of detecting sophisticated APTs with high accuracy and efficiency. By leveraging autoencoders, Spatial-Convolutional GNNs, Memory-based models, and federated learning with encrypted weights, we address the critical challenges in modern cybersecurity, providing a comprehensive and scalable solution for detecting APTs.

\section{Related Work}
\label{sec:related-work}
In this section we provide a comprehensive background and literature review of recent works in various areas relevant to our research. We focus on several key aspects: GNN-based APT detection, graph-data generation, decentralised and distributed IDS architectures, and encryption techniques in federated learning. This review highlights the existing methodologies and their limitations, setting the stage for the novel contributions of our study.
\subsection{GNNs in Cybersecurity}
Graph Neural Networks have emerged as a highly effective tool for cybersecurity, particularly in the domain of intrusion detection. Traditional Intrusion Detection Systems often rely on rule-based or signature-based approaches \citep{han2020unicorn}, which struggle to adapt to rapidly evolving attack strategies and novel threats. In contrast, GNNs offer the ability to model complex relationships within network structures, enabling them to identify subtle, concealed-patterns commonly associated with APTs.

\begin{figure}[h]
\centering
    \begin{minipage}{0.27\textwidth}
    \centering
        \includegraphics[width=0.98\textwidth]{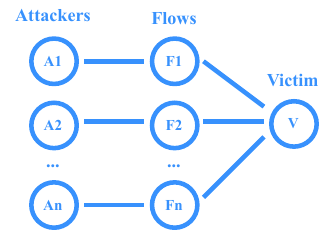}

        \vspace{2mm}
        (a) DDOS
        
    \end{minipage}
    \medskip
    \begin{minipage}{0.27\textwidth}
        \centering
        \includegraphics[width=0.98\textwidth]{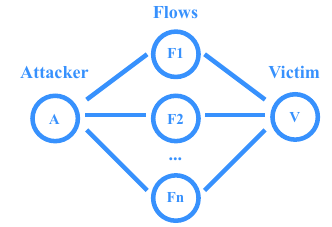}

        \vspace{2mm}
        (b) Port Scan
    \end{minipage}
        \begin{minipage}{0.27\textwidth}
    \centering
        \includegraphics[width=0.98\textwidth]{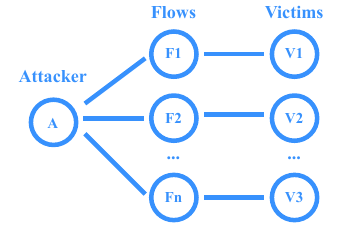}

        \vspace{2mm}
        (c) Network Scan
    \end{minipage}
    \medskip
    \caption{Graph signature of well-known attacks \citep{pujol2022unveiling}}
\label{fig:nids}
\end{figure}

 By leveraging the ability to capture intricate interconnections between network entities, GNNs can provide a more robust detection mechanism compared to conventional IDS. For instance, \cite{pujol2022unveiling} demonstrated GNNs' capability in anomaly detection by capturing network-flow relationships, offering precise identification of malicious activities such as Distributed Denial of Service (DDoS), port scans, and network scans. Examples of graph modelisation of attacks is shown in Figure \ref{fig:nids}.

Recent studies further emphasize the effectiveness of GNNs in cybersecurity. \cite{wang2022graph} highlighted how GNNs can tackle the challenges of applying deep learning to structured graph data, showing their utility across various domains, including social networks and bioinformatics. The mathematical framework of GNNs aligns well with real-world network structures, making them an ideal choice for anomaly detection. Additionally, \cite{bilot2023graph} explored the application of GNNs for detecting APTs, underscoring their potential in enhancing IDS systems by identifying sophisticated threat patterns.

While GNNs offer a transformative approach to intrusion detection by modeling complex network relationships, their effectiveness relies on the availability of high-quality datasets and careful handling of challenges such as data biases and concept drift \citep{tsymbal2004problem}. Furthermore, implementing GNNs in cybersecurity requires significant technical expertise to fully exploit their capabilities.

While these datasets provide valuable resources for evaluating PIDSes and graph-based APT detection models, there is still a noticeable gap in publicly available benchmarks specifically tailored for graph detection of APTs, as highlighted by \citep{liu2023causal}. Many datasets are either outdated or insufficiently diverse to capture the complexity of modern APT campaigns. Furthermore, the confidential nature of APT attack data restricts the availability of real-world datasets, making it challenging for researchers to apply machine learning and graph-based techniques effectively. Continued development of diverse, high-quality datasets is essential to advancing the field of APT detection.

\subsection{Graph Generation}
Graph data, such as provenance graphs used in intrusion detection, can be tailored based on the specific requirements of each model \citep{zhong2024survey}. Depending on the nature of the task, models may operate on either static or dynamic graph data, with the latter being essential for real-time analysis, such as flow-based network monitoring \citep{kazemi2020representation}.
\subsubsection{Static-data}
Static graphs, with their fixed topology and connections, represent the network or system at a specific moment in time. These types of graphs are commonly used in graph-based IDS, as seen in studies by \citep{protogerou2021graph, lo2022graphsage}, and \cite{wang2023n}. Before being input into models, static graph data can be modified through techniques such as enrichment or transformation to improve the model’s ability to detect anomalies and cyber threats.

Enriching static graphs involves adding more information to the original graph to enhance its structure. For example, in the work of \cite{zheng2019gcn}, additional edges and hyper-edges were introduced, such as connecting two-hop neighbors, to capture more complex relationships within the data. Similarly, \cite{cheng2021discovering} used methods like one-hot encoding and similarity measures to extract new attributes for nodes and edges, allowing the model to gain deeper insights into the graph’s structure and enhance detection accuracy.

Another approach to modifying static graphs is through transformation, where the graph structure is altered to serve different analytical purposes. \cite{chang2021graph, zhu2022graph} explored transforming original graphs into line graphs, where each vertex corresponds to an edge in the original graph. This transformation can be particularly useful for converting tasks like edge classification into node classification, streamlining the analysis and allowing for more efficient detection of malicious activities, as discussed by \cite{zhong2024survey}.

\subsubsection{Dynamic-data}
Provenance graphs are generally-speaking static data, i.e, they represent the network or system in a fixed topology. However, in a real-time intrusion detection, data is dynamic, i.e represented in flows or sequences processed separately over time. Dynamic methods involves mainly two techniques, snapshoting and sketching \citep{zhong2024survey}.

Snapshoting approach involves dividing the data into intervals and creating a snapshot for each interval. Each snapshot represents the state of the network during a specific time period, allowing for the analysis of temporal changes and patterns. For example, \cite{king2023euler, jedh2021detection} proposed techniques that use specific time units or intervals as snapshots to analyze the network's structure and relationships over time. \cite{xiao2021learning} extended this approach by considering each timestamp as a node, thereby constructing a graph with a fixed number of timestamps at each instance. This method effectively captures temporal dynamics in the network data.

Differently, sketching techniques involves creating a summarized or approximate representation of the graph to reduce complexity while retaining essential features. This method focuses on efficiently capturing the essential information from a large and complex network. \cite{paudel2020snapsketch} introduced a graph-sketching technique that converts a graph into a low-dimensional sketch vector using a simplified hashing technique. \cite{messai2023iot} proposed constructing an activity graph from networking events during a monitoring period. This activity graph captures both structural and semantic features from network traffic, which are then used to train a neural network to distinguish between normal activities and attacks.

\subsection{GNNs in APT attacks Detection}
The utilisation of GNNs in Intrusion Detection Systems has been at the forefront of recent advancements in Cybersecurity, particularly for detecting APT attacks. Various GNN-based architectures have been developed to enhance the detection of these sophisticated threats by leveraging the inherent graph structure of system data. For instance, MAGIC \citep{jia2023magic} employs a masked-graph auto-encoder to learn meaningful graph representations. This technique involves masking certain nodes or edges during training to reconstruct these masked elements, allowing the model to understand deep correlations within the provenance graphs. By using multiple Graph Attention Network \citep{velivckovic2017graph} layers as a decoder in the self-supervised learning process, MAGIC \citep{jia2023magic} predicts the original graph-structure from corrupted versions. This approach enables the model to capture outliers at both the system-entity and batched-log levels, facilitating multi-granularity detection. It can precisely identify the sub-graph containing the attacks and the specific nodes involved, enhancing the accuracy of APT detection. 
Another significant work, KAIROS, developed by \cite{cheng2023kairos}, introduces a novel GNN-based encoder-decoder architecture that focuses on learning the temporal evolution of a graph and assessing the degree of anomalousness in data at different timestamps. When KAIROS \citep{cheng2023kairos} detects a new edge in the provenance graph, it processes the feature vectors of the source node's neighbors using a Temporal Graph Network \citep{rossi2020temporal} encoder. This encoder generates an embedding for the new edge, which is subsequently reconstructed using a Multilayer Perceptron (MLP) \citep{popescu2009multilayer} decoder. This architecture not only aids in understanding benign system behavior, but also facilitates the efficient detection of suspicious nodes. Furthermore, KAIROS \citep{cheng2023kairos} tracks output edges from these nodes to construct a summary graph, enabling system administrators to backtrack and analyze attack scenarios.
Graph Competitive Autoencoder (GCA), proposed by \cite{ye2023detect}, is another GNN-based work that utilizes a Graph Convolutional Network (GCN) \citep{kipf2016semi} to discover attack scenarios through intrusion alert-correlation. This architecture uses a GCN as the graph encoder to generate embeddings, converting each graph into a vector representation that captures features of both attack and benign graphs. GCA \cite{ye2023detect} employs separate decoders for labeled benign graphs and unlabeled graphs, enhancing classification accuracy. Attack graphs are identified by comparing reconstruction errors; if the attack decoder’s error is lower than the benign decoder’s for an unlabeled graph, it is classified as an attack graph.
Another research, named GHUNTER, depicted by \cite{cheng2023ghunter}, aims to enhance the discriminatory capability of Graph Isomorphism Network \citep{xu2018powerful} and their generalisation of Weisfeiler-Lehman (WL) heuristic \citep{leman1968reduction} for subgraph matching in APT detection. GHUNTER \cite{cheng2023ghunter} operates by using a simple GIN \citep{xu2018powerful} to generate node embeddings from provenance graphs. It then seeks a node whose embedding indicates that this node serves as an anchor for a subgraph resembling a previously known malicious graph. This approach bolsters the model’s ability to detect subgraphs that match known attack patterns. 
Further more, two other researches explored the implementation of GraphSAGE \citep{hamilton2017inductive} in the detection of APTs; the first one is Threatrace, developed by \cite{wang2022threatrace}, it employs GraphSAGE to embed evolving graphs for timely threat detection. This architecture efficiently learns node roles in provenance graphs despite not considering every node’s neighbor. By focusing on the temporal dynamics of network activities, Threatrace \citep{wang2022threatrace} is capable of detecting anomalies and suspicious activities that span multiple stages and time periods, providing a comprehensive view of potential APT campaigns. The second one is XFedGraph-Hunter, explored by \cite{son2023xfedgraph}, represents a pioneering effort in applying Federated Learning to PIDS applications using GraphSAGE for APT detection. This architecture employs a pre-trained transformer to add edge features and standardize node features before GraphSAGE layers (TGraphSage). It also uses GNNExplainer \citep{ying2019gnnexplainer} to generate explainable detection graphs. By leveraging FL, XFedGraph-Hunter \citep{son2023xfedgraph} demonstrated improved precision and F1-score compared to the original GraphSAGE model. The collaborative training on local data, where each host sends weight updates to a server for aggregation, enabled better generalization and reduced false-positive rates.

These GNN-based IDS architectures highlight the innovative approaches to APT detection, leveraging the capabilities of GNNs to model complex relationships and patterns within graph data. These systems provide powerful tools for cybersecurity professionals to enhance the security of network systems against advanced cyber threats. However, a key limitation of these approaches is the high rate of false positives that often accompanies GNN-based detection models \citep{altaf2024gnn, cheng2023kairos, jia2023magic}. This issue arises when benign activities that share similarities with attack patterns are incorrectly flagged as malicious, resulting in alert fatigue and reduced efficiency of security teams \citep{cheng2023kairos}. Additionally, the reliance on pre-constructed, static graphs in many of these approaches limits their ability to detect dynamic, time-sensitive attack stages that evolve over time. As a result, static models often miss critical attack behaviors unfolding in real-time, particularly in long-term APT campaigns.

\subsection{Decentralised and Distributed APT Detection}
Few works only have studied the application of decentralised or distributed IDS architecture, especially in the detection of APT attacks. Most APT-detectors \citep{jia2023magic, cheng2023kairos, wang2022threatrace, cheng2023ghunter} rely on centralized data analysis, which can be resource-intensive and pose significant privacy risks. 

In a pioneer work, \cite{wu2022paradise} introduced a distributed IDS architecture named Paradise, which utilizes Kafka servers \citep{thein2014apache} to distribute client data to IDS servers for parallel attack detection. Although this architecture improves resource efficiency and enables parallel processing, it also faces challenges related to bandwidth consumption and data privacy. By transmitting system-logs and other sensitive information over the network, this approach is vulnerable to sniffing attacks .

In response to these challenges, federated learning offers a more secure alternative. By allowing clients to train local models and share only model parameters with the central server, FL facilitates the development of models capable of identifying anomalous behavior indicative of potential intrusions while ensuring that sensitive data remains within the local environment of each client, which minimizes the risk of data exposure and reduces network strain. For instance, \cite{son2023xfedgraph} explored the application of FL on GNNs for APT attacks detection and demonstrated that FL could reduce the false-positive rate and enable the model to generalize to unknown attacks through collaborative model-training.

\subsection{Encryption in Federated Learning}
While FL improves data privacy by keeping raw data local, it still involves sharing model updates, which can be intercepted by malicious actors whom can reconstruct the clients' data using reverse engineering \citep{shanmugarasa2023systematic}. This vulnerability has led to the exploration of various encryption techniques to secure the communication of model weights.

In our previous research, called FedHE-Graph \citep{mansour2024fedhe}, we addressed some of these challenges in APT attacks detection by testing federated learning with hybrid encryption, using Advanced Encryption Standard (AES) for the symmetric encryption \citep{dworkin2001advanced}, and Rivest-Shamir-Adleman (RSA) for the asymmetric one \citep{rivest1978method} . This approach demonstrated promising results by reducing execution time and enhancing privacy. However, it showed limitations against inference attacks \citep{krumm2007inference}, where an attacker targets the server during the decryption phase to intercept the model weights. Intercepting this information enables attackers to conduct Mimicry Attacks, mimicking the usual behavior of a user to evade detection by the IDS \citep{goyal2023sometimes}. Attackers may also leverage intercepted data for information gathering or for conducting poisoning attacks to degrade the quality of model training-data, posing significant risks \citep{yang2017generative}.

Homomorphic encryption \citep{yi2014homomorphic} offers a solution to this issue by allowing computations on encrypted data without the need for decryption. This ensures that even if an attacker gains access to the server, they cannot access the model weights. To the best of our knowledge, no IDS has explored the use of Homomorphic encryption in the detection of APT attacks. Yet, one significant contribution in this field is the work by \cite{chen2019multi}, which introduced Multi-key Homomorphic Encryption (MKHE) for federated learning. In this scheme, each client generates a pair of private/public keys for encrypting and decrypting model weights. The server can perform computations on the encrypted weights without decrypting them, thus maintaining data confidentiality throughout the learning process. This method enhances the security of APT detection systems by safeguarding against various attack-vectors.
This research has shown that integrating Homomorphic Encryption with FL can effectively protect against security threats, including MitM attacks and server-side breaches. However, the computational overhead introduced by encryption can be substantial, necessitating efficient implementation-strategies to balance security and performance.

\section{Our approach : CONTINUUM}
\label{sec:our-approach}
In this section, we propose a novel architecture for Host-based Intrusion Detection Systems utilizing a heterogeneous-graph approach. This architecture, which we called \emph{Continuum}, efficiently captures complex-dependencies between system entities to detect APT attacks. Designed for scalability, Continuum can easily adapt to new hosts, integrate additional data streams, and optimize resource usage. It addresses current GNN model limitations, offering a new solution for protecting organizations against APT threats.

\begin{figure}[h]
\centering
\includegraphics[width=0.8\textwidth]{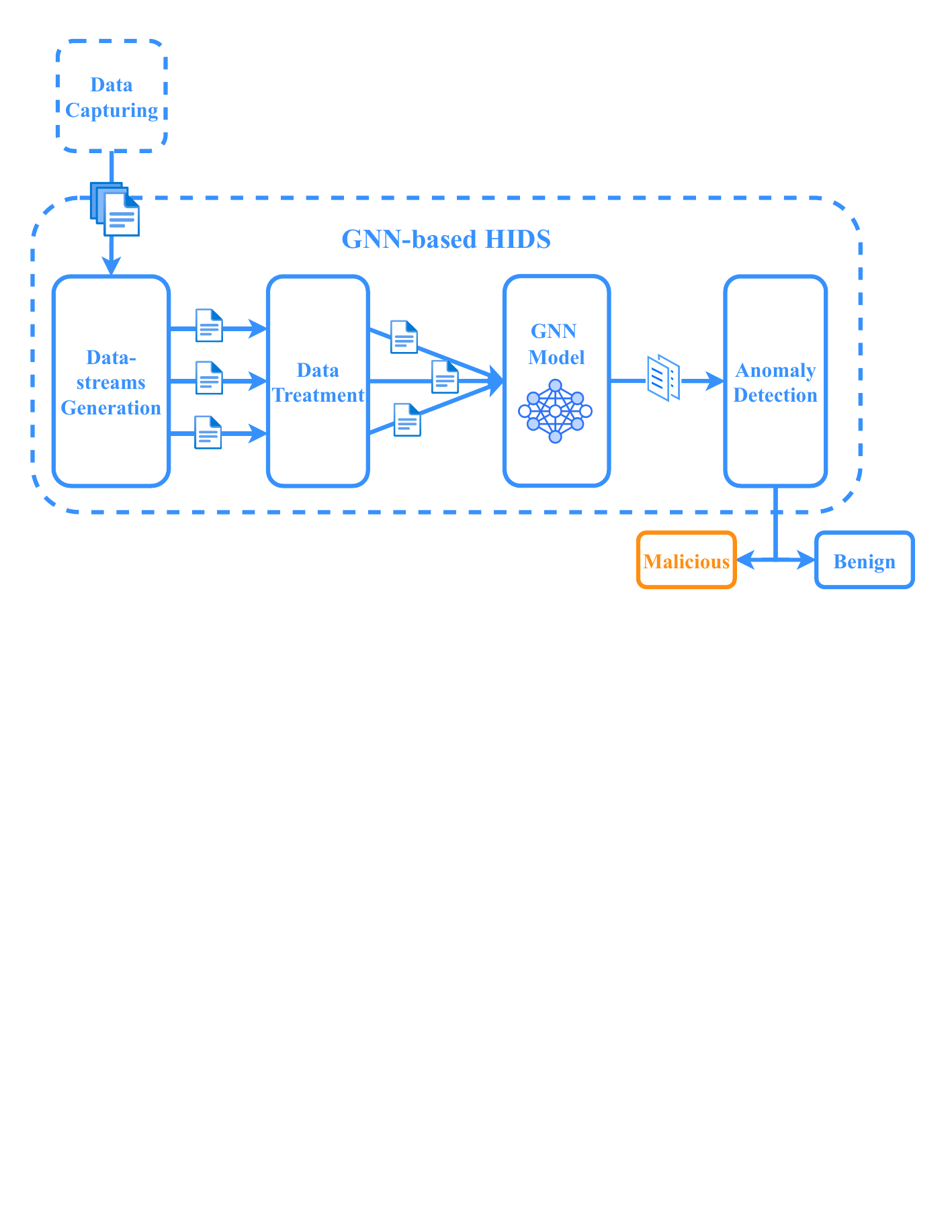}
\vspace{-9cm}
\caption{Global architecture of Continuum}
\label{fig:architecture}
\end{figure}

The architecture is depicted in Figure \ref{fig:architecture}. It begins with provenance data captured by a third-party system data-flow analyzer. This data includes the relationships and dependencies between various system entities such as files, scripts, users, and processes. The provenance data is structured as graphs, representing system interactions that will serve as the input for both training and detection phases.

After capturing the data, it undergoes pre-processing to convert it into data streams, also referred to as snapshots, that represent specific periods of system activity. These snapshots are subject to quality enhancement processes to ensure that the data is ready for use in the GNN model. The GNN model then processes these snapshots to learn complex dependencies between the system entities, producing node embeddings that capture detailed relationships within the system.

As for the training, the system is pre-trained exclusively on benign data, which enables the GNN model to learn what constitutes normal behavior within the host. As a result, the model can more effectively detect anomalies, which may indicate the presence of an APT attack. By analyzing these embeddings in real time, the model can differentiate between benign and malicious activities, forming the core of the APT detection mechanism.

Moreover, Continuum operates by detecting APTs at both the graph level (entire snapshots) and entity level (individual nodes). During real-time operation, incoming data-streams are processed into snapshots and fed into the GNN. The model evaluates each snapshot sequentially, identifying any anomalies that deviate from normal behavior, thus flagging potential APTs.

This design ensures proactive detection by analyzing real-time provenance data, providing a scalable, GNN-based approach to safeguard systems from sophisticated APT attacks.
\subsection{Model construction}
The Continuum architecture is designed to leverage a Spatial-Temporal Graph Neural Network to detect APTs by capturing both spatial relationships and temporal dynamics between system entities. This section details each component of the model, justifying the design choices made to ensure robust APT detection.
\subsubsection{STGNN}
\label{subsubsec:stgnn}
A key trend in APT detection is the use of autoencoders instead of traditional classifiers in PIDS architectures \citep{ye2023detect, cheng2023kairos, jia2023magic}. In an autoencoder-based IDS, the encoder produces embeddings for graph nodes, while the decoder minimizes the similarity errors between actual and reconstructed node-representations. This process preserves essential information and original attack-patterns. Based on this, we choose to incorporate a GNN-based autoencoder in our design. Autoencoders are crucial for uncovering complex-patterns related to APT attacks by generating precise-embeddings for nodes and extracting hidden-information from their neighbors.

\begin{figure}[h]
\centering
\includegraphics[width=0.9\textwidth]{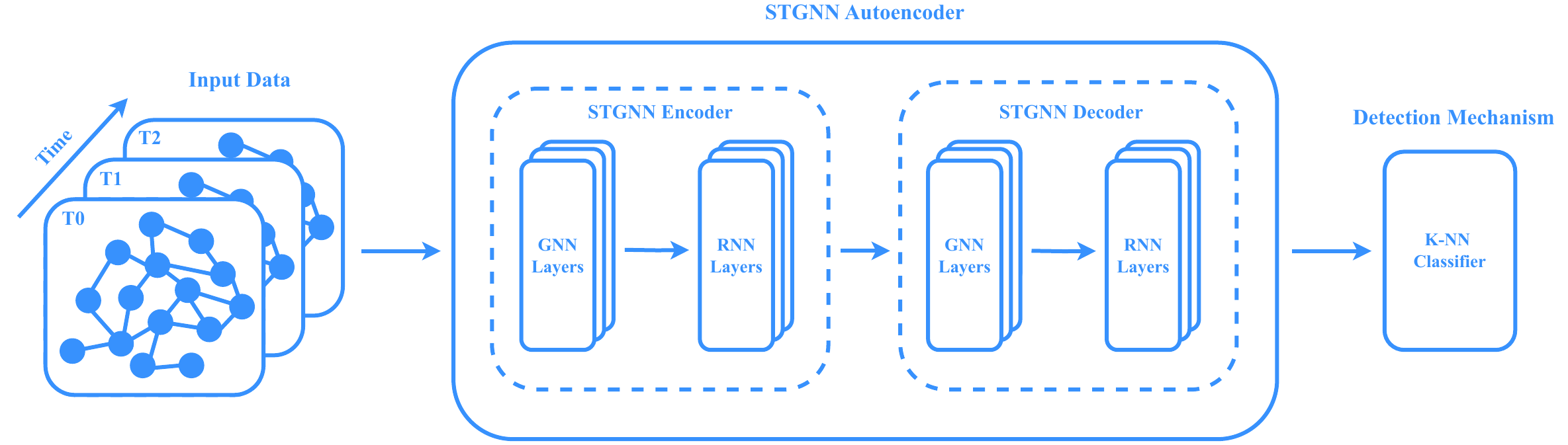}
\caption{Architecture of our GNN autoencoder}
\label{fig:autoencoder}
\end{figure}

Moreover, detecting APT attacks relies on two main characteristics: identifying abnormal behaviors indicative of an attack, and recognizing that APT patterns evolve over time. Therefore, our IDS must consider the temporal dimension of APT attacks across various stages. To address these aspects, we implement an STGNN in our autoencoder, as shown in Figure \ref{fig:autoencoder}. STGNN is chosen for its ability to model the complex and dynamic nature of APT attacks. The STGNN integrates spatial information (how entities interact within a single snapshot of the system) and temporal information (how these interactions evolve over time). This dual modeling allows the system to detect both immediate, anomalous activities, and prolonged malicious-behaviors that unfold gradually.

The choice of STGNN over traditional GNNs is essential because APTs often involve multistage attacks that are not confined to a single timeframe. The temporal dimension enables the system to track the progression of the attack over time, which is critical in identifying threats that do not exhibit immediate anomalous behavior.
\paragraph{Spatial Information}
Graph-data aggregates information from nodes and their neighborhoods, making it beneficial to use a GNN model that emphasizes a localized view of each node’s surroundings. This is especially relevant for Spatial Convolutional GNNs, which utilize message-passing functions to uncover hidden information from neighboring nodes and learn complex data correlations. Their application in APT detection is strategically sound, as APT attacks often reveal complex behaviors through interactions within local neighborhoods rather than isolated nodes. SpGNNs excel in capturing these intricate relationships by focusing on local connectivity patterns \citep{sahili2023spatio}. By focusing on the immediate neighbors of each node, the system can capture fine-grained anomalies, such as unexpected file executions or unauthorized access to processes. This localized information is then aggregated over multiple GNN layers, building a richer representation of each node’s role within the system. Notable SpGNN architectures, such as GraphSAGE and GAT, are recognized for their effectiveness in learning from local graph structures \citep{jia2023magic, son2023xfedgraph, wang2022threatrace}. GraphSAGE aggregates information from node neighborhoods, while GATs dynamically weigh node contributions using attention mechanisms. Thus, SpGNNs are crucial for enhancing IDS capabilities against sophisticated cyber threats like APT attacks.
\paragraph{Temporal Information}
Given the prolonged and stealthy nature of APT attacks, analyzing the temporal evolution of provenance data is key to detecting anomalies at different stages of an attack. As attackers move through distinct phases over time, monitoring the relationships between system entities across various periods helps reveal evolving attack patterns. To achieve this, a memory-based model, such as an RNN, GRU, or LSTM, is employed in Continuum. The RNN layers track the temporal dependencies of each system entity by maintaining a hidden state that updates as new snapshots are processed. This hidden state serves as memory for the system, enabling it to identify slow-developing attacks that might evade detection in single-snapshot models. Moreover, unlike conventional feedforward neural networks \citep{bebis1994feed}, which treat inputs independently, memory-based models leverage their internal state to process sequential information. This allows them to predict behaviors based on temporal relationships.

The use of RNNs is critical for modeling temporal dependencies and attack patterns. Many APT attacks involve a sequence of seemingly unrelated activities that, when viewed in isolation, appear benign. By leveraging the temporal dimension, the system can correlate these activities, identifying a coherent attack sequence that might otherwise go unnoticed.
\subsubsection{Encoder}
The encoder plays a key role in transforming raw graph-data into meaningful representations. It consists of multiple layers of GNNs that produce node-embeddings dense vector representations that encode each node's local context (i.e., relationships with its neighbors) \citep{jia2023magic}. These embeddings are passed through the RNN layers, which enrich them with temporal information. This process ensures that the final embeddings reflect not only the current state of each entity but also how its behavior has evolved over time. 
\begin{equation}
\mathbf{node\_embeddings} = \text{RNN}\left(\left[\text{GNN}(G_t)\right]_{t=1}^{T}\right)
\end{equation}
where 
 \( G_t \) is the embedding of graph \( G \) at timestamp \( t \),
  \( GNN \) is the update function of the GNN layer, and 
 \( RNN \) is the update function of the RNN layer.
The choice of this multi-layered encoder structure is justified by the need for a hierarchical understanding of system behavior. The GNN layers focus on spatial aggregation, while the RNN layers add a temporal perspective, resulting in embeddings that are highly informative and tailored for anomaly detection in APT scenarios, which are known for being persistent and prolonged in time.

\subsubsection{Decoder}
Once the embeddings are generated, the decoder attempts to reconstruct the node features and interactions, ensuring that the embeddings accurately represent the original graph-structure \citep{jia2023magic}.
This is particularly important in an anomaly detection context, where even slight deviations from normal behavior need to be identified. The better the embeddings represent normal system activity, the more sensitive the system will be to subtle APT-related anomalies.

Our decoder utilizes an STGNN for this purpose, creating an objective function that boosts the graph representation module's performance. It regenerates initial node-embeddings, allowing for the calculation of feature-reconstruction loss, which helps enhance the relevance of the generated embeddings.

The reconstruction loss is computed using a loss function that compares the initial node vectors with the reconstructed ones, aiming to maximize the behavioral information in the abstracted node embeddings.
\begin{equation}
\mathcal{L}(I, R) = \mathbf{loss\_function}(I, R)
\end{equation}
where
\( {L} \) is the Reconstruction loss,
\( I \) are the initial node vectors, and 
 \( R \) are the reconstructed node vectors.
Additionally, incorporating RNN layers into the decoder preserves temporal information by merging the current snapshot's output with the upcoming one. This continuity captures temporal dependencies between snapshots, improving the decoder's ability to reconstruct node features and refining the model’s overall performance.
\subsubsection{Detection Mechanism}
In GNN-based IDS, attacks are detected through two primary approaches: classification and anomaly detection.

The classification approach assigns predefined labels to entities based on patterns learned from benign behaviors. For example, \cite{manzoor2016fast} simulated machine states by modeling activities such as web browsing, watching YouTube, or playing video games. A classifier compares new entities to these known behaviors and assigns them to one of the predefined classes. If no match is found, the entity is labeled as malicious.

In contrast, anomaly detection identifies deviations from normal behavior without relying on predefined class labels, this is the case of some recent GNN-based APT detectors \citep{cheng2023kairos, jia2023magic}. This approach functions as a binary classification, flagging any significant departure from typical behavior as potentially malicious.

Both classification and anomaly detection are crucial for detecting and addressing threats in network environments, providing complementary perspectives on security monitoring and threat detection. In Continuum, we opt for an anomaly-detection model, which classifies nodes as either benign or anomalous, by using a K-Nearest Neighbors (K-NN) algorithm \citep{peterson2009k}. 
K-NN is chosen for its flexibility in identifying novel attack patterns. APTs are constantly evolving, and attackers frequently modify their tactics to evade detection. By comparing the learned embeddings of each node to those of known benign entities, the system can identify outliers—nodes whose behavior deviates from the norm. This allows the system to adapt to new and unforeseen threats, making it highly effective in real-world environments where APTs are constantly evolving

During model exploitation in Continuum, the node embeddings generated by the STGNN encoder will be used for either node or graph classification. For graph classification, the graph encoder applies average pooling to create a graph embedding from the node embeddings produced by the final RNN layer.
\begin{equation}
    \mathbf{H_g} = \frac{1}{|V|} \sum_{n \in V} \mathbf{H_n}
\end{equation}
where
\( H_g \) is the embedding of graph \( G \),
\( H_n \) is the embedding of node \( n \) of the last RNN, and
 \( V \) is the set of neighboring nodes \( n \).

Finally, the benign embeddings are clustered using the K-NN model, serving as reference points in classifying instances as either malicious or benign based on the distance to their nearest neighbors. 

\noindent The detailed architecture of the STGNN is depicted in Figure \ref{fig:autoencoder_detailed} below.

\begin{figure}[h]
\centering
\includegraphics[width=0.85\textwidth]{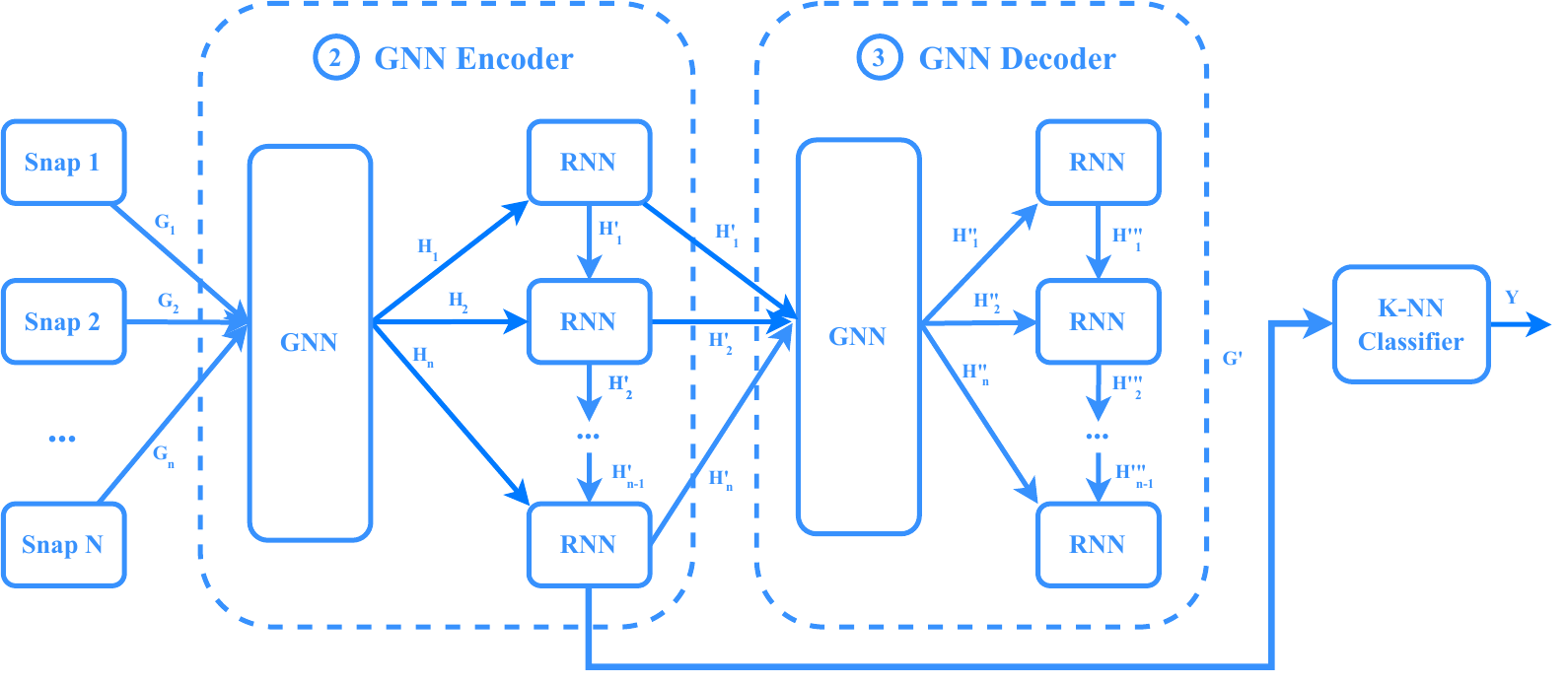}
\caption{Detailed architecture of our GNN autoencoder}
\label{fig:autoencoder_detailed}
\end{figure}

\subsection{Data construction}
Another component of our solution involves a data-to-graph methodology centered on provenance graphs, the Graph Neural Network model is trained using data snapshots derived from these graphs. However, current APT datasets in provenance graph format are typically in log form and lack segmentation into snapshots \citep{manzoor2016fast, DVN/8GKEON_2018, DVN/IA8UOS_2018, DVN/69SMQB_2020, DVN/KUNDIU_2020, githubGitHubDarpai2oTransparentComputing, alsaheel2021atlas}.

To address this, we developed a pre-processing method that uses timestamps, one-hot encoding, and edge compression. Timestamps segment the dataset into distinct intervals, creating snapshots, while one-hot encoding is applied to standardize the representation of different edge and node types. Additionally, edge compression reduces redundancy, optimizing the dataset size and improving model-training efficiency. This pre-processing method is crucial in preparing APT datasets for effective use in our GNN-based detection framework, as it adapts to various graph-based datasets, enhancing performance in cybersecurity applications.
\subsubsection{Snapshots generation}
In the context of APT detection, dataset splitting plays a critical role in facilitating the temporal analysis of events. Each row in APT datasets typically represents system interactions as nodes and edges in a graph \citep{manzoor2016fast}, accompanied by metadata such as timestamps. These timestamps indicate the specific moments when events were captured, providing chronological insight into system activities.

\begin{figure}[h]
\centering
\includegraphics[width=0.8\textwidth]{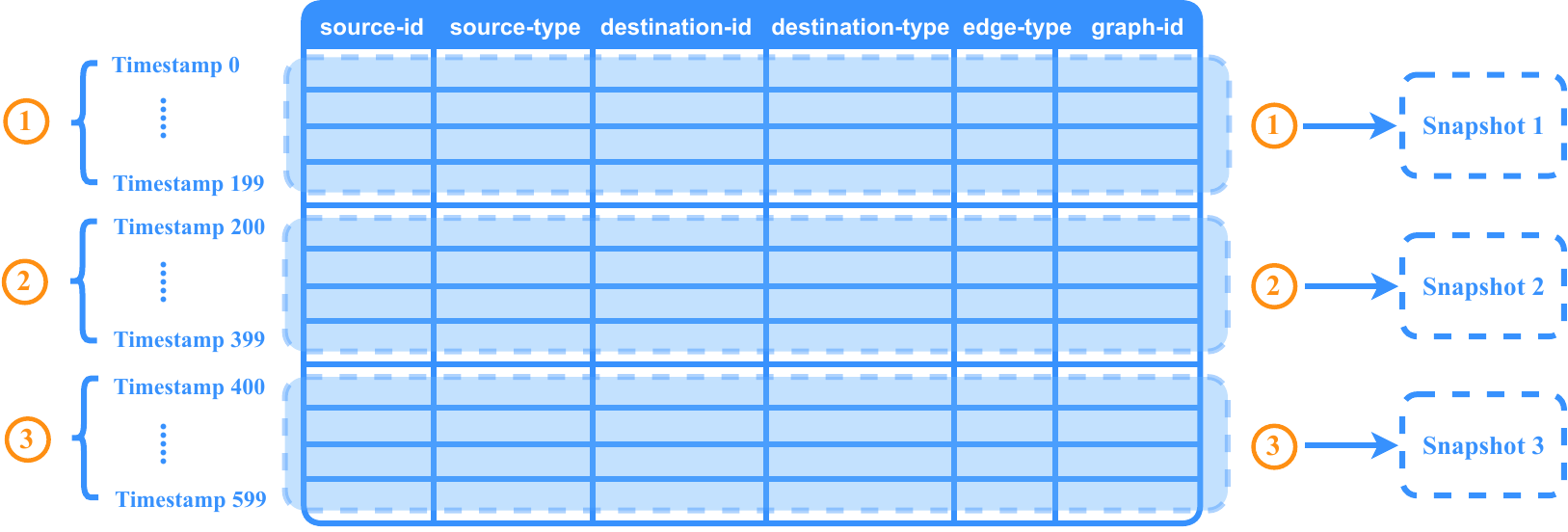}
\caption{Dataset splitting in snapshots}
\label{fig:dataset-split}
\end{figure}

To effectively process and analyze APT datasets, we leverage this chronological information by segmenting the datasets into discrete time-intervals based on the timestamps of the events. This process of timestamp-based segmentation allows us to split the data into distinct sets of events, each corresponding to a specific time-range. For instance, we depict in Figure \ref{fig:dataset-split} a simplified example-dataset with 600 events occurring at various timestamps. We can divide the timestamps into three distinct-intervals. The first interval may cover events from timestamp 0 to 199, the second from 200 to 399, and the final interval includes the remaining events. These segmented sets are referred to as snapshots, and they enable a more granular analysis of the dataset, focusing on the evolution of events over time.

The first phase of pre-processing involves extracting nodes, edges, and their associated attributes from the dataset. Each line in the dataset represents a specific action or event, which varies in format depending on the dataset in use. The process is summarized in Algorithm \ref{algo:data-preparation}, where each line is processed individually to extract relevant information such as node types, edge types, and timestamps. 
\begin{algorithm}
  \KwIn{Dataset file}
  \KwOut{Prepared file}
  - Initialize the graph\;
  \For{\textup{each line in the dataset-file} }{
    - Separate the fields of the line using a delimiter \;
    - Store each field in a specific variable \;
    - Create the nodes of the edge if not already present in the graph with their types\;
    - Create the edge in the graph with its type and its timestamp.\;
    - Append the new line to the output file \;
  }
  - Close the input file \;
  \Return the graph\;
  \caption{Information extraction from datasets}
  \label{algo:data-preparation}
\end{algorithm}

The snapshot generation process follows the steps outlined in Algorithm \ref{algo:data-snapshoting}.

\begin{algorithm}
  \KwIn{Formatted dataset, Number of snapshots}
  \KwOut{Graphs, Node dimension, Edge dimension}
  \For{\textup{each graph (file) of the dataset} }{
    - Extract node/edge types, timestamps, and maximum timestamp from the graph \;
    - Extract node/edge dimension which is the number of node/edge types \;
    - Divide the maximum timestamp by the number of snapshots to calculate the length of the time sequence of each snapshot  \;
    \For{\textup{each snapshot} }{
        - Generate a directed graph \;
        - Append all the nodes to the graph \;
        - Append only the edges appearing in the time-sequence with the same index as the snapshot \;
        - Save the snapshot in PKL format \;
  }
  }
  \Return the snapshots and their node/edge dimensions \;
  \caption{Snapshot division from datasets}
  \label{algo:data-snapshoting}
\end{algorithm}

By organizing the dataset into these snapshots, we not only streamline the pre-processing phase but also gain the ability to observe and study temporal patterns, correlations, and relationships within the dataset more effectively. This structured approach enhances our ability to detect and analyze APT attacks that evolve in stages over time, as it provides a clear temporal-framework for the behavior of system entities.

\subsubsection{One-Hot encoding}
To effectively manipulate the provenance graphs within each snapshot, it is crucial to encode both the nodes and edges into a standardized format. This encoding process enables the system to process the diverse types of nodes and edges found in APT datasets, facilitating their use in the subsequent stages of GNN processing. In our framework, we employ one-hot encoding for both nodes and edges to represent the various types of interactions and entities captured within the system.
\paragraph{Node encoding}
Each node in the provenance graph represents a system entity such as a process, file, or network socket. Since the dataset may include different types of nodes depending on the source, we assign a one-hot encoded vector to represent each node type. For example, consider a graph with the following types of nodes:
\begin{equation*}
V_{i} = (Process_i, File_i, Socket_i)
\end{equation*}
Each node $v_k$ in this set is encoded as a one-hot vector where each entry corresponds to a node type. The encoding is as follows:
\begin{equation}
v_k = \left. \begin{cases}
1, & \text{if node type of $v_{i}$ matches type $k$} \\
0, & \text{otherwise}
\end{cases} \right. ; \forall k \in [0, 2]
\end{equation}
This method ensures that each node type is uniquely represented, enabling the system to differentiate between processes, files, and sockets. By applying one-hot encoding, the dataset is transformed into a structured format that GNN models can efficiently interpret, allowing for the identification of entity behaviors within the graph.
\paragraph{Edge encoding}
\begin{figure}[h]
\centering
    \begin{minipage}{0.6\textwidth}
    \centering
        \includegraphics[width=\textwidth]{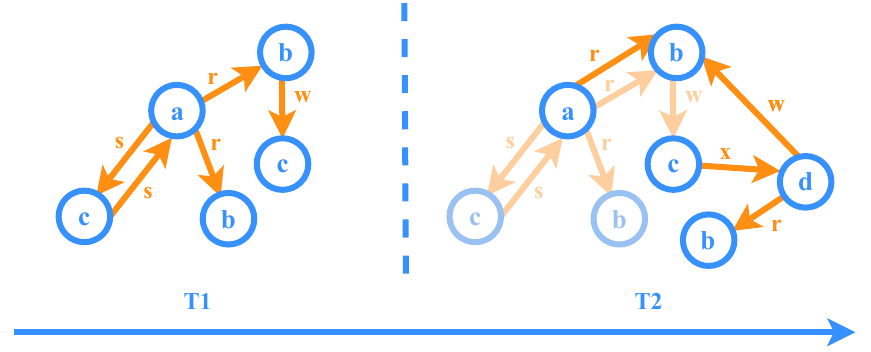}

        \vspace{2mm}
        (a) Graph representation of two snapshots before encoding
        \vspace{4mm}
    \end{minipage}
    \medskip
    \begin{minipage}{0.6\textwidth}
        \centering
        \includegraphics[width=\textwidth]{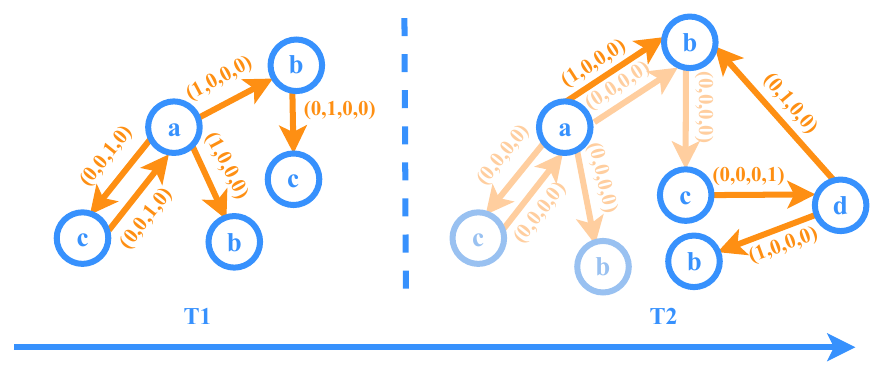}

        \vspace{2mm}
        (b) Graph representation of two snapshots after encoding
    \end{minipage}
    \caption{Example of edge encoding for two snapshots. The actions represented are Read (r), Write (w), Send (s), and Execute (x).}
\label{fig:edge-encoding}
\end{figure}

Similar to nodes, the edges in the provenance graph are also encoded using a one-hot encoding scheme. Each edge represents an interaction between two nodes, such as a process reading from a file, or a network socket sending data. Since datasets contain various types of edges corresponding to different system-actions, it is important to standardize these representations to ensure consistency across different datasets.

For example, we encode the following four types of actions: Read, Write, Send, and Execute. The edge 
$e_{ij}$ between two nodes $(i, j)$ is represented by a tuple $E_{ij}$ as follows:
\begin{equation*}
E_{ij} = (Read_{ij}, Write_{ij}, Send_{ij}, Exec_{ij})
\end{equation*}
Each component $e_k \in E_{ij}$ is encoded as:
\begin{equation}
e_k = \left. \begin{cases}
1, & \text{if edge type of $e_{ij}$ matches action $k$} \\
0, & \text{otherwise}
\end{cases} \right. ; \forall k \in [0, 3]
\end{equation}
This standardized-encoding approach ensures that the provenance graph maintains consistency across datasets and provides the foundation for subsequent dataset-compression. The compression process optimizes the data for storage and computational efficiency without sacrificing the richness of the graph's structure.

By using these one-hot encoding techniques for both nodes and edges, the dataset becomes ready for GNN-based analysis, enabling efficient-processing and manipulation of the provenance graphs.

An example of edge encoding is depicted in Figure \ref{fig:edge-encoding}. In this scenario, we simulate a toy-dataset prolonged over two snapshots, each containing a specific number and type of edges. Edges of each snapshot are encoded separately. Note that an edge appearing in the first snapshot should not appear in the second one as each snapshot describes a set of events happening in a fixed interval of time. However, we choose to represent them in the referred figure for illustration only. The information represented on edges of the first snapshot is transmitted to the second snapshot through the memory of the RNN-layers presented in Figure \ref{fig:autoencoder}.

\subsubsection{Edge compression}
After encoding the dataset into the new format, we optimize its size through compression by combining edges with identical source and destination nodes within each snapshot. These edges are summed into a single compressed-edge, as represented by:
\begin{equation}
C_{ij} = \sum E_{ij} = (\sum Read_{ij}, \sum Write_{ij}, \sum Send_{ij}, \sum Exec_{ij})
\end{equation}
where  
\( C_{ij} \) is the compressed edge between the pair of nodes \( (i, j) \), and 
 \( E_{ij} \) is the encoded edge between the pair of nodes \( (i, j) \).
To illustrate this method, we show an example of compression on a simple graph in Figure \ref{fig:compression} below. 

\begin{figure}[h]
\centering
\includegraphics[width=0.6\textwidth]{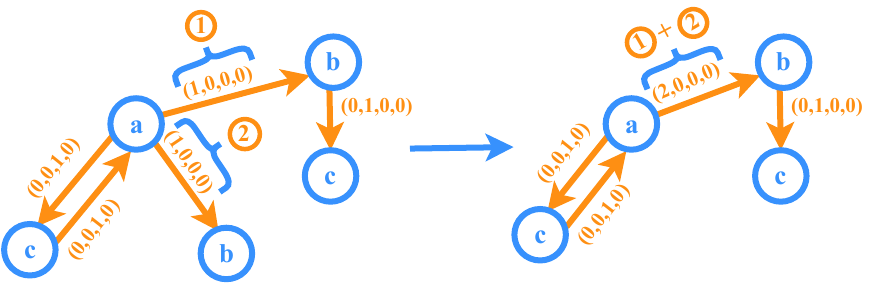}
\caption{Compression of two edges}
\label{fig:compression}
\end{figure}

\begin{figure}[h]
\centering
    \begin{minipage}{0.8\textwidth}
        \centering
        \includegraphics[width=\textwidth]{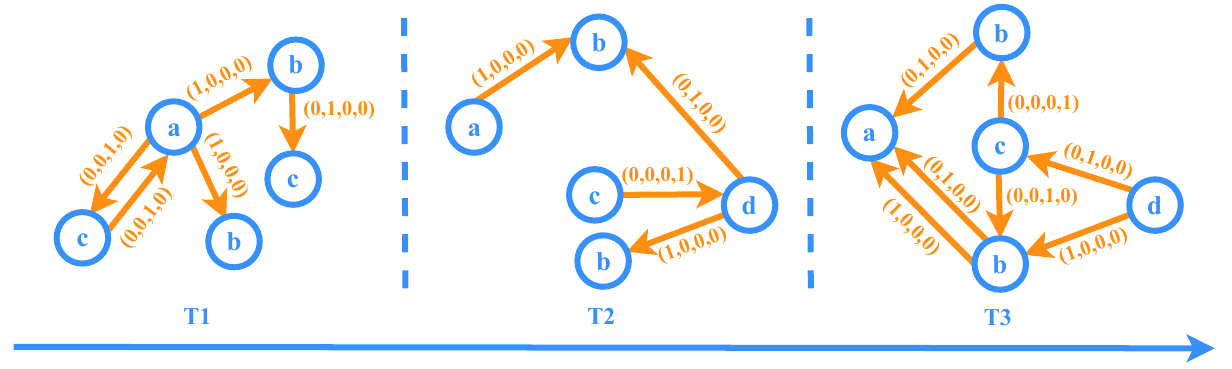}

        \vspace{2mm}
        (a) Graph representation of three snapshots before compression
        \vspace{4mm}
    \end{minipage}
    \medskip
    \begin{minipage}{0.8\textwidth}
        \centering
        \includegraphics[width=\textwidth]{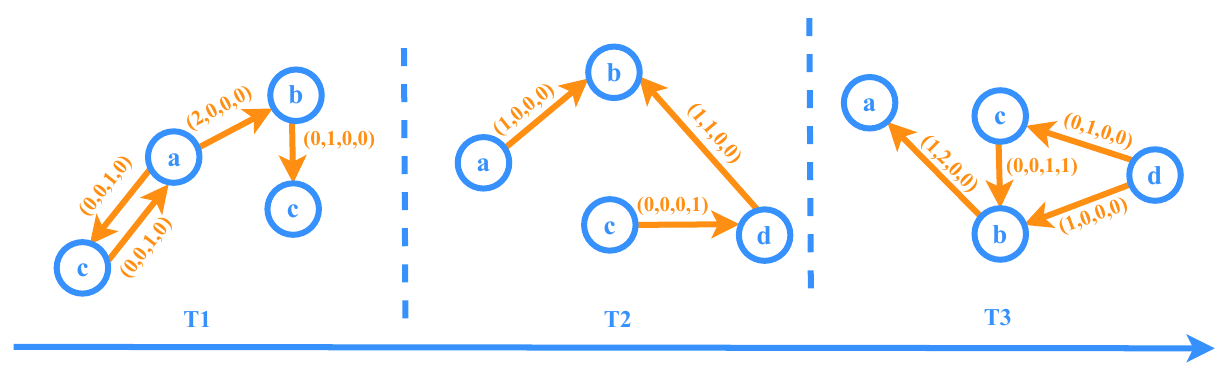}

        \vspace{2mm}
        (b) Graph representation of three snapshots after compression
    \end{minipage}
    \caption{Example of edge compression for three snapshots.}
\label{fig:edge-compression}
\end{figure}

For further explanations, we depict in Figure \ref{fig:edge-compression} the application of the compression methods on three snapshots generated from the previous edge-encoding scenario (Refer to Figure \ref{fig:edge-encoding}).

We notice that this method reduces the number of edges in the dataset, effectively minimizing its size, which can often be in the tens of gigabytes. To confirm this claim, we show in Table \ref{tab:dataset-compression} a size comparison between datasets before and after compressing their edges. 

For instance, the size of Theia-E3 dataset \citep{githubGitHubDarpai2oTransparentComputing} was reduced by over 3.7 million edge, and Clearscope-E3 \citep{githubGitHubDarpai2oTransparentComputing} saw a reduction of over 2.3 million edges.
This comparison reveals an average 70.8\% reduction of edges in general, which means a significant reduction in datasets complexity, training times, resource, and energy consumption, which directly improves the model-training efficiency.

\begin{table}[h]
\centering
\caption{Number of edges per dataset before and after compression}
\label{tab:dataset-compression}
\vspace{2mm}

\begin{tabular}{c c c c}
\hline
\multirow{2}{*}{\textbf{Detection mechanism}} & \multirow{2}{*}{ \textbf{Dataset}} & \multirow{2}{*}{\textbf{\#Edges before compression}} & \multirow{2}{*}{\textbf{\#Edges after compression}} \cr \\
\hline

\multirow{6}{*}{\textbf{Graph-level}} & \textbf{Cadets-Unic } & {5 625} & {5 625} \\ 
\cline{2-4}
& \textbf{Clrscope-E3} & {2 344 070} & {15 240}  \\ 
\cline{2-4}
& \textbf{SC-2} & {917 608} & {715 338}  \\ 
\cline{2-4}
& \textbf{Streamspot} & {149 618} & {13 190} \\ 
\cline{2-4}
& \textbf{Wget} & {148 982} & {99 171} \\ 
\cline{2-4}
& \textbf{Wget-HL} & {106 738} & {83 134} \\ 
\hline

\multirow{3}{*}{\textbf{Node-level}} & \textbf{Cadets-E3} & {1 648 006} & {840 299} \\ 
\cline{2-4}
& \textbf{Theia-E3} & {4 319 197} & {574 964} \\ 
\cline{2-4}
& \textbf{Trace-E3} & {1 178 021} & {811 205}  \\ 
\hline

\end{tabular}
\end{table}
The compression process is outlined in Algorithm \ref{algo:data-compressing}.

At the end of the compression process, the compressed snapshots are transformed into \texttt{Deep Graph Library (DGL)} graphs using the \texttt{dgl.from\_networkx} method, which prepares them for input into the GNN model. \texttt{DGL} offers built-in GNN architectures and graph-structured data classes that are essential for implementing machine-learning tasks on graph data.

\begin{algorithm}
  \KwIn{Snapshot, Nodes dimension, Edges dimension}
  \KwOut{Graph}
  - Append the nodes of the snapshot to the new graph \;
  \For{\textup{each edge of the snapshot} }{
    - Extract the attributes of the edge \;
    \eIf{\textup{the source and destination nodes of this edge already have an edge in the new graph}}{
        - Add the edge type of the new edge to the existing one \;
    }{
    - Append the edge to the new graph \;
    }
  }
  \Return the new graph \;
  \caption{Edge compression in snapshots}
  \label{algo:data-compressing}
\end{algorithm}


\subsection{Model deployment}
Establishing an efficient model-deployment strategy is critical for optimizing APT-detection performance. Traditional approaches using local models and centralized deployments \citep{han2020unicorn, wang2022threatrace, yan2022deepro, ye2023detect, cheng2023kairos, jia2023magic} are resource-intensive and may hinder Intrusion Detection Systems' adaptability to evolving APT threats. Some efforts, like the Paradise framework \citep{wu2022paradise}, have explored distributed deployment using Kafka servers \citep{thein2014apache}. Paradise distributes client data to IDS servers via load-balancing, which alleviates resource strain on hosts; but is bandwidth-intensive, and exposes client data to potential privacy risks, such as sniffing attacks.

Federated Learning offers a more efficient alternative by enabling clients to train models on their benign-data locally and share only model-parameters with the server. This approach reduces network strain and the risk of data leaks, as only model -weights are exchanged. FL allows the server to aggregate these parameters, avoiding the need to train separate models for each client. Similar approaches, like XFedGraph-Hunter \citep{son2023xfedgraph}, have applied FL to Graph Neural Networks for APT detection, reducing false positives and improving model-generalization to unknown attacks.

However, FL is still vulnerable to data leaks through Reverse Engineering \citep{shanmugarasa2023systematic}, where a Man-in-the-Middle attacker could intercept model weights and reconstruct the original data. To address this, we propose an FL-based deployment strategy that includes encrypting model-weights before transmission. This ensures secure communication and client privacy, preventing MitM attacks while optimizing network-resources. This strategy allows collaborative learning of benign behaviors across the system, addressing the third hypothesis.
\subsubsection{Large-scale deployment}
Federated learning within Intrusion Detection Systems enables the creation of models that generalize well to new data through collaborative learning and knowledge-sharing. This allows the generated model to adapt to benign host-behaviors observed during training and integrate newly-identified APT attack patterns from participating hosts. Knowledge exchange occurs through the sharing of model parameters between the central server and client hosts.

By aggregating model-parameters across various networks, the system facilitates extensive knowledge-sharing on APT attacks. For example, organizations within the same industry or research labs in a region can contribute their local server weights to develop a more generalized model. This process involves sharing encrypted-model weights with a central server for aggregation, using homomorphic encryption to ensure the confidentiality of each organization's model details from the external server.

This method promotes a collaborative cybersecurity-alliance among organizations, improving the overall knowledge base of the Graph Neural Network model. Leveraging this collaborative approach enhances the precision and efficiency of APT detection, offering critical support in an era of increased targeting of companies and governments by threat actors.
\subsubsection{Encryption methodology}
Based on the literature review presented earlier, our IDS must be resilient against both Reverse Engineering and Inference attacks to prevent Man-in-the-Middle attacks aimed at stealing client data or injecting noisy data into the server \citep{krumm2007inference}. Such attacks can introduce biases during the server's weight aggregation, impairing the model's ability to differentiate between benign and malicious data, thus increasing False Positive and False Negative rates.

\begin{figure}[h]
\centering
\includegraphics[width=0.7\textwidth]{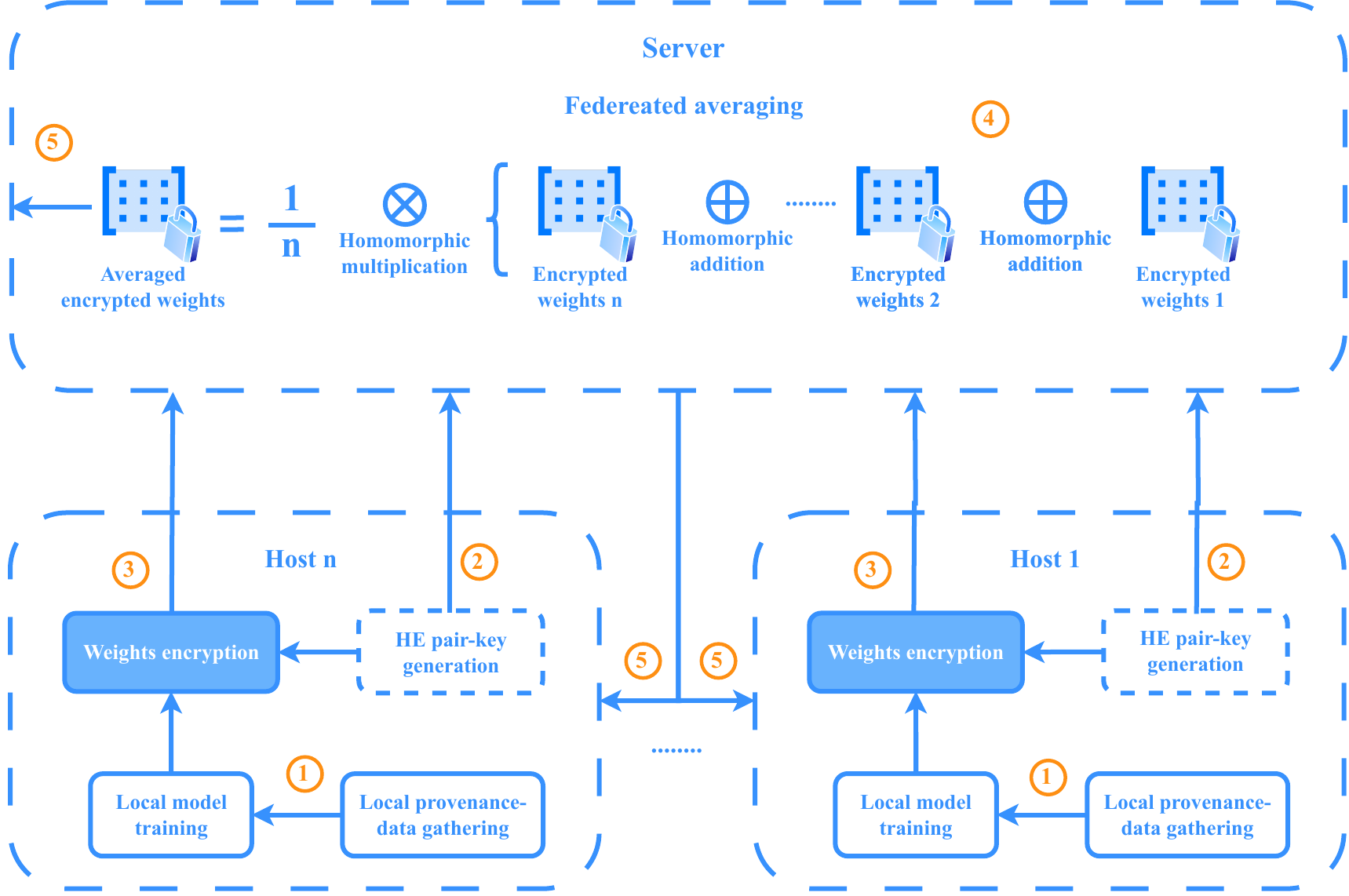}
\caption{Homomorphic encryption in Continuum}
\label{fig:homomorphFedhe}
\end{figure}

To counter these risks, the server must be treated as an untrusted entity, preventing access to client data. This is achieved through Homomorphic Encryption \citep{yi2014homomorphic}, a method proven effective in Federated Learning within sensitive environments like hospital management systems \citep{viensea1106}. Similarly, in APT detection, clients' data reflects machine and user behaviors, making it valuable to attackers who can use it for Mimicry Attacks, evading IDS detection \citep{goyal2023sometimes}, or for data poisoning to degrade model quality \citep{yang2017generative}. A general overview of the deployment architecture is shown in Figure \ref{fig:homomorphFedhe}.

More technically, we opt for implementing Multi-key Homomorphic Encryption \citep{chen2019multi}, where each client generates a private/public key pair. The private key encrypts and decrypts model weights, while the server uses public keys to compute on encrypted weights without decrypting them, thus protecting data from malicious actors. This encryption, deployed within FL, safeguards against Mimicry Attacks, data interception, and other security threats, reinforcing the IDS's defenses in APT detection.

We integrated Homomorphic Encryption into our federated learning system using \texttt{FedML-HE} \citep{jin2023fedml}, which employs a threshold key management scheme based on Shamir’s secret sharing \citep{shamir1979share}. In this approach, a trusted key authority generates a secret key, which is split into shares distributed among the clients. Each client uses its key-share to encrypt local model-weights before sending them to the server through an \texttt{MQTT} or \texttt{MPI} communication.

The server aggregates the encrypted weights using a \emph{Somewhat-Homomorphic Encryption} system\footnote{Somewhat homomorphic is a homomorphic encryption-system that supports both addition and multiplication, but only a limited number of times due to noise accumulation}, allowing operations on encrypted data without access to the underlying values. The aggregated ciphertext is then sent back to the clients, where each client partially decrypts the weights using its key-share. The clients exchange their partially-decrypted weights and collaboratively reconstruct the fully-decrypted weights in plaintext.

The main steps of this process are outlined in Algorithm \ref{algo:threshold-management}.

\begin{algorithm}
  \KwIn{Aggregator (server), N clients}
  \KwOut{Plaintext weights}
    \tcp{Key Generation}
  - Key Authority generates a secret key $K$ divided into $N$ shares $k_{1},k_{2},...,k_{N}$ \;
  \tcp{Weights Encryption}
  \For{\textup{each client $i$ of $N$}}{
    - Encrypt the local-parameters $W_{i}$ using the key-share $k_{i}$ \;
    - Send the encrypted parameters $W_{i}$ to the server \;
    }
    \tcp{Aggregation Phase}
    - The server aggregates the encrypted-parameters $W_{1}, W_{2},...,W_{n}$ \;
    - The server sends the aggregated encrypted-parameters $W$ to the clients \;
    \tcp{Decryption Phase}
    \For{\textup{each client $i$ of $N$}}{
        - Partially decrypt W using $k_{i}$ \;
        - Send the partially-decrypted parameters $W'_{i}$ to the other clients \;
        - Wait to receive the decrypted-parameters from all the other clients \;
        - Merge the decrypted-weights $W'_{1},W'_{2},...,W'_{n}$\;
        }
  \Return the plaintext weights \; 

  \caption{FedML-HE with threshold-key management}
  \label{algo:threshold-management}
\end{algorithm}

\section{Performance evaluation}
\label{sec:Perf}
In this section, we 
focus onto the concrete implementation of the CONTINUUM framework. We 
 detail the technical aspects of the system, specifically the development of the components within the Continuum architecture. 
 We cover mainly the development of the GNN model, focusing on the training, validation, and evaluation stages. 
\subsection{Test environment}
We implemented our GNN-based IDS -Continuum- using \texttt{Python}, leveraging popular libraries for machine learning, graph manipulation, and federated learning, including \texttt{FedML}\footnote{\url{https://www.fedml.ai/}} for federated-learning deployment, and \texttt{TenSEAL} \citep{Benaissa2021TenSEALAL} for encryption. These libraries were instrumental in simplifying the complexities of both data processing and model deployment, while ensuring that the system remained secure and scalable.

For the model training, we utilized a virtual machine (VM) equipped with NVIDIA Graphics Processing Units (GPUs), and the \texttt{CUDA}\footnote{\url{https://developer.nvidia.com/cuda-toolkit}} platform. \texttt{CUDA}, a parallel-computing platform developed by NVIDIA, enabled significant acceleration in the training process by offloading heavy computations to the GPU. This GPU-accelerated training environment was essential in handling the large-scale datasets and computationally-intensive tasks involved in graph-based machine learning. \texttt{CUDA}’s efficient parallel-processing dramatically reduced the time required for model training compared to traditional Central Processing Unit (CPU)-based training
 particularly when working with complex, deep learning architectures.


To ensure compatibility and stability across different dependencies, we used \texttt{Python} version 3.10.14 along with the latest versions of key libraries in Machine Learning and Graph Learning.

All the experiments were conducted on a machine equipped with 16GB of RAM, a 12GB VRAM GPU (RTX 3060), and a 6-core CPU (i5-12400F). These resources are well-suited for training on large APT datasets, although lighter resources would suffice for smaller datasets.

%
%
%
%
  
\subsection{Datasets}
Various provenance-based datasets for APT attacks detection have been introduced in the literature, with some serving as benchmarks and others remaining relatively-unexplored. In this work, we select datasets from both categories to train and evaluate our solution. This approach allows us to compare our results with state-of-the-art IDS benchmarks while also promoting new datasets for future research in the field.

The datasets chosen for our evaluation include Wget \citep{DVN/IA8UOS_2018}, Wget-HourLong \citep{DVN/8GKEON_2018}, SC-2 \citep{han2020unicorn}, Streamspot \citep{manzoor2016fast}, and the E3-Darpa datasets, which consist of Cadets, Clearscope, Trace, and Theia \citep{githubGitHubDarpai2oTransparentComputing}. These datasets exhibit different characteristics in terms of size, the number of nodes, edges, and graphs they contain, as summarized in Table \ref{tab:dataset-sizes}. This comparison highlights the variation in complexity and diversity across the datasets.

\begin{table}[h]
\centering
\caption{Datasets description}
\label{tab:dataset-sizes}
\vspace{2mm}

\begin{tabular}{c c c c c c}
\hline
\multirow{2}{*}{\textbf{Granularity}} & \multirow{2}{*}{\textbf{Dataset}} & \multirow{2}{*}{\textbf{Scenario}} & \multirow{2}{*}{\textbf{Avg. \#Graphs}} & \multirow{2}{*}{\textbf{Avg. \#Nodes}} & \multirow{2}{*}{\textbf{Avg. \#Edges}} \cr\\

\hline
\multirow{16}{*}{\textbf{Graph-level}} &
\multirow{2}{*}{\textbf{Wget}} 
& Benign & 125 & 37,296 & 148,982 \\
\cline{3-6}
& & Attack & 25 & 39,215 & 152,879  \\
\cline{2-6}

& \multirow{2}{*}{\textbf{Wget-HourLong}} 
& Benign & 100 & 28,664 & 107,589   \\
\cline{3-6}
& & Attack & 5 & 29,384 & 104,438   \\
\cline{2-6}

& \multirow{2}{*}{\textbf{Clearscope E3}} 
& Benign & 44 & 2,307 & 2,097,957   \\
\cline{3-6}
& & Attack & 50 & 11,972 & 2,560,650   \\
\cline{2-6}

& \multirow{2}{*}{\textbf{Cadets-Unicorn }}
& Benign & 110 & 2,704 & 3,017   \\
\cline{3-6}
& & Attack & 3 & 56,303 & 100,474   \\
\cline{2-6}

& \multirow{2}{*}{\textbf{SC-2}} 
& Benign & 125 & 238,338 & 911,153   \\
\cline{3-6}
& & Attack & 25 & 243,658 & 949,887  \\
\cline{2-6}

& \multirow{2}{*}{\textbf{Streamspot}} 
& CNN & 100 & 8,989 & 294,903 \\
\cline{3-6}
& & Download & 100 & 8,830 & 310,814\\
\cline{3-6}
& & Gmail & 100 & 6,826 & 37,382  \\
\cline{3-6}
& & VGame & 100 & 8,636 & 112,958  \\
\cline{3-6}
& & Youtube & 100 & 8,292 & 113,229 \\
\cline{3-6}
& & Attack & 100 & 8,890 & 28,423  \\

\hline
\multirow{8}{*}{\textbf{Node-level}} & \multirow{2}{*}{\textbf{Theia E3}} 
& Benign & - & 1,598,647 & \multirow[c]{2}{*}{2,874,821}  \\
\cline{3-5}
& & Attack & - & 25,319 & \\
\cline{2-5}

& \multirow{2}{*}{\textbf{Cadets E3}} 
& Benign & - & 1,614,189 & \multirow[c]{2}{*}{3,303,264} \\
\cline{3-5}
& & Attack & - & 12,846 & \\
\cline{2-6}

& \multirow{2}{*}{\textbf{Trace E3}} 
& Benign & - & 3,220,594 & \multirow[c]{4}{*}{4,080,457} \\
\cline{3-5}
& & Ext. Backdoor & - & 732 &  \\
\cline{3-5}
& & Pine Backdoor & - & 67,345 &  \\
\cline{3-5}
& & Phishing Exe. & - & 5 & \\
\hline
\end{tabular}
\end{table}

Despite their differences, the datasets share common types and attributes related to the nodes and edges, providing insights into the communication patterns and data transfers among system entities. A summary of these commonalities is presented in Table \ref{tab:common-infos}, offering a clearer understanding of the similarities and differences in dataset structures.

\begin{table}[t]
    \centering
    \caption{Common information among APT datasets}
    \label{tab:common-infos}
    \vspace{2mm}
    
    \begin{tabular}{c c c}%
        \hline
        \multirow{2}{*}{\textbf{Subject}} & \multirow{2}{*}{\textbf{Common Types}} & \multirow{2}{*}{\textbf{Common Attributes}} \cr \\
        
        \hline 
        \multirow{2}{*}{\textbf{Nodes}} & \multirow{2}{*}{File, Process, Socket, Task} & \multirow{2}{*}{Path, PID, TTY, TIME\_CPU, Node Type} \cr \\
        
        \hline
        \multirow{2}{*}{\textbf{Edges}} & \multirow{2}{*}{Read, Write, Execute, Send, Receive, Open, Close, Clone} & \multirow{2}{*}{Src/Dst Node, Edge Type, Timestamp} \cr \\
   
        \hline
    \end{tabular}
    
\end{table}

\subsection{Model construction}
Our overall solution is built around a Spatial-Temporal Graph Auto-encoder, as outlined in the system design (Refer to Section \ref{subsubsec:stgnn}). This section details the key choices made for the GNN and RNN layers within our architecture, explaining how these components contribute to the detection of Advanced Persistent Threats through both spatial and temporal data-modeling.

In order to find the best combination of layers, we explored multiple spatial-convolutional GNNs, including \texttt{GATConv}, \texttt{GCNConv}, \texttt{GINConv}, and \texttt{SageConv} from the \texttt{dgl.nn} library in \texttt{Python}, alongside various RNN layers such as \texttt{GRUCell}, \texttt{RNNCell}, and \texttt{LSTMCell} from \texttt{torch.nn}. The k-NN layer was implemented using the \texttt{kneighbors} method from \texttt{sklearn.neighbors.NearestNeighbors}.

After extensive hyperparameter tuning and empirical evaluation, 
we identified Graph Attention Networks \citep{velivckovic2017graph} as the optimal choice for capturing spatial-dependencies, and Gated Recurrent Units for temporal-sequence modeling. However, instead of the standard GAT implementation from \texttt{dgl.nn}, we adopted the GAT layer from Magic \citep{jia2023magic}, which integrates edge features into the attention mechanism. Unlike the additive attention in the \texttt{dgl.nn.GATConv} class, where the attention weight $\alpha$ is computed as:
\begin{equation}
\alpha = Droupout\left(Softmax\left(LeakyReLU\left(Linear\left(feat_{src} | feat_{dst}\right)\right)\right)\right)
\end{equation}
Magic incorporates edge features, enhancing the expressiveness of the model by adjusting the attention weight $\alpha$ as:
\begin{equation}
\alpha = Droupout\left(Softmax\left(LeakyReLU\left(Linear\left(feat_{src} | feat_{edge} | feat_{dst}\right)\right)\right)\right)
\end{equation}
This architectural choice improves the model's ability to capture both node and edge-level interactions, proving critical for detecting complex patterns in the context of APT attacks.

\noindent \textbf{\underline{Training phase: }} The training of our model is conducted using a supervised-learning approach, employing either graph-level or entity-level methods. In both cases, a batch-based strategy is adopted, where the model iterates over multiple epochs, processing snapshots of a single benign-graph at a time.

Each batch consists of the entire benign-graph, allowing the loss function to account for the reconstruction features of all nodes simultaneously. The training begins by activating the training mode via \texttt{model.train()} and resetting gradients using \texttt{optimizer.zero\_grad()}. For each benign graph, we compute the loss by comparing the original node features with the reconstructed features, leveraging symmetric binary cross-entropy \citep{wang2019symmetric} for accuracy. The implementation provided by \cite{jia2023magic} is employed for this loss function.

Model weight-updates are done through back-propagation using \texttt{loss.backward()} and \texttt{optimizer.step()}. We utilized the \textbf{Adam} optimizer (\texttt{torch.optim.Adam}) for efficient gradient-based optimization, and employed the \texttt{PreLu} activation function in the GNN layers. The model was trained for 50 epochs in the case of entity-level training and 6 epochs for graph-level training, reflecting optimal settings derived from our experimental analysis.

\noindent \textbf{\underline{Validation phase: }} The validation process plays a key role in tuning hyper-parameters during training. In the graph-level approach, graph embeddings are derived from node embeddings generated by the encoder, using the \texttt{torch.mean()} function to aggregate them. For the entity-level approach, we directly utilize the node embeddings without further aggregation.

To evaluate the model's performance, we cluster the training embeddings by employing the \texttt{sklearn.neighbors. \newline NearestNeighbors} method, and compute the mean distance between each benign point and its neighbors using \texttt{kneighbors()}. This establishes a baseline for comparison during validation.

For the validation set, embeddings are generated using the STGNN encoder, which are then fed into the pre-trained k-NN model. We calculate the mean distances between each validation point and its neighbors in the same manner as for the training set.

The validation score is computed by dividing the mean distance of validation points by that of the training points. Performance is evaluated through precision-recall and Receiver Operating Characteristic (ROC)-Area Under the Curve (AUC) metrics, calculated respectively using the \texttt{sklearn.metrics.precision\_recall\_curve} and the \texttt{sklearn.metrics.roc\_auc\_score} methods. The optimal threshold for classifying test points as benign or malicious is determined based on these metrics.

\noindent \textbf{\underline{Evaluation phase: }} Similar to the validation phase, the evaluation begins by generating clusters of benign data-points. The classification of test data-points is based on their proximity to the nearest neighbors within these benign clusters. Specifically, a test point is classified as malicious or benign by measuring its distance from its nearest benign-neighbors, following the same methodology applied during validation.


\noindent \textbf{\underline{Integration of Federated Learning: }} We implemented Federated Learning in our GNN model using \texttt{FedGraphNN} benchmark \citep{he2021fedgraphnn} from \texttt{FedML} to avoid building the FL system from scratch. \texttt{FedML} provides built-in functionality to configure a federated server and its clients, handling communication via, either\texttt{Message Queuing Telemetry Transport (MQTT)} for real-world scenarios, or \texttt{Message Passing Interface (MPI)} for simulations.

\subsection{Metrics}
To accurately evaluate the performance of our model, we employed four key metrics: Precision, Recall, F1-Score, and AUC. These metrics offer insights into different aspects of the model’s classification performance, especially when dealing with imbalanced datasets, which is typical in APT detection. The metrics are calculated based on the following classification outcomes:
\begin{itemize}
    \item [] \textbf{True Positives (TP):} Malicious samples (APT attacks) correctly identified.
    \item [] \textbf{True Negatives (TN):} Benign samples correctly identified.
    \item [] \textbf{False Positives (FP):} Benign samples incorrectly classified as malicious.
    \item [] \textbf{False Negatives (FN):} Malicious samples incorrectly classified as benign.
\end{itemize}

Our focus was primarily on optimizing Precision and F1-Score during hyper-parameter tuning, as our goal was to minimize false negatives to catch undetected attacks and reduce false positives to limit false alarms.
\begin{itemize}
    \item [] \textbf{Precision:} The ratio of correctly-predicted positive observations to the total predicted-positives. High precision indicates a low false-positive-rate (FPR), making it crucial for minimizing false alarms.
    \begin{equation}
        Precision = \frac{TP}{TP + FP}
    \end{equation}
    \item [] \textbf{Recall:} The ratio of correctly-predicted positive observations to all actual positive-observations. High recall ensures that the model captures most-relevant results, though it may increase false positives.
    \begin{equation}
        Recall = \frac{TP}{TP + FN}
    \end{equation}
    \item [] \textbf{AUC:} Used to measure the quality of the model by evaluating the area under the ROC curve, which plots the true-positive rate (TPR) against the FPR across different threshold values. A higher AUC score indicates better performance in distinguishing between classes.
    \begin{equation}
        AUC \approx \sum \left(\text{width of $FPR$ interval} * \text{average height of $TPR$}\right)
    \end{equation}
    \item [] \textbf{F1-Score:} The harmonic mean of precision and recall, providing a balanced measure that accounts for both FPs and FNs. A higher F1-Score signifies a better trade-off between precision and recall.
    \begin{equation}
        F1-Score = 2 * \frac{Precision * Recall}{Precision + Recall} 
    \end{equation}
\end{itemize}

These metrics provide a comprehensive evaluation of our model, helping us gauge its ability to detect APT attacks while minimizing misclassifications.

\subsection{Results and comparison}
We begin by showcasing the results obtained from our Graph Neural Network (GNN)-based model across all datasets, considering two distinct configurations: without Federated Learning and with Federated Learning. These results are summarized in Table \ref{tab:fl--results-comparison}.

\begin{table}[h]
\centering
\caption{Best performances of our solution on different datasets}
\label{tab:fl--results-comparison}
\vspace{2mm}

\begin{tabular}{c c c c c c c}%
\hline
\multirow{2}{*}{\textbf{Detection}} & \multirow{2}{*}{\textbf{Dataset}} & \multirow{2}{*}{\textbf{Solution}} & \multirow{2}{*}{\textbf{Precision}} & \multirow{2}{*}{\textbf{Recall}} & \multirow{2}{*}{\textbf{F1-Score}} & \multirow{2}{*}{\textbf{AUC}} \cr \\
\hline

\multirow{12}{*}{\textbf{Graph-level}} & \multirow{2}{*}{\textbf{Cadets-Unicorn}} & Without FL & {{1.0}} & {{1.0}} & {{1.0}} & {{1.0}} \\
\cline{3-7}
& & \cellcolor{blue!12}{\textbf{With FL}} & \cellcolor{blue!12}{\textbf{1.0}} & \cellcolor{blue!12}{\textbf{1.0}} & \cellcolor{blue!12}{\textbf{1.0}} & \cellcolor{blue!12}{\textbf{1.0}}  \\
\cline{2-7}
& \multirow{2}{*}{\textbf{Clearscope-E3}} & Without FL & {1.0} & {1.0} & {1.0} & {1.0} \\
\cline{3-7}
& & \cellcolor{blue!12}{\textbf{With FL}} & \cellcolor{blue!12}{\textbf{1.0}} & \cellcolor{blue!12}{\textbf{1.0}} & \cellcolor{blue!12}{\textbf{1.0}} & \cellcolor{blue!12}{\textbf{1.0}}  \\
\cline{2-7}
& \multirow{2}{*}{\textbf{SC-2}} & \textbf{Without FL} & \textbf{0.85} & \textbf{0.92} & \textbf{0.88} & \textbf{0.92} \\
\cline{3-7}
& & \cellcolor{blue!12}{\textbf{With FL}} & \cellcolor{blue!12}{\textbf{0.77}} & \cellcolor{blue!12}{\textbf{0.96}} & \cellcolor{blue!12}{\textbf{0.86}} & \cellcolor{blue!12}{\textbf{0.88}} \\
\cline{2-7}
& \multirow{2}{*}{\textbf{Streamspot}} & Without FL & {1.0} & {1.0} & {1.0} & {1.0} \\
\cline{3-7}
& & \cellcolor{blue!12}{\textbf{With FL}}& \cellcolor{blue!12}{\textbf{1.0}} & \cellcolor{blue!12}{\textbf{1.0}} & \cellcolor{blue!12}{\textbf{1.0}} & \cellcolor{blue!12}{\textbf{1.0}} \\
\cline{2-7}
& \multirow{2}{*}{\textbf{Wget}} &\textbf{Without FL} & \textbf{1.0} & \textbf{1.0} & \textbf{1.0} & \textbf{1.0} \\
\cline{3-7}
& & \cellcolor{blue!12}{\textbf{With FL}} & \cellcolor{blue!12}{\textbf{1.0}} & \cellcolor{blue!12}{\textbf{0.96}} & \cellcolor{blue!12}{\textbf{0.98}} & \cellcolor{blue!12}{\textbf{0.97}} \\
\cline{2-7}
& \multirow{2}{*}{\textbf{Wget-HourLong}} & Without FL & {0.83} & {1.0} & {0.91} & {0.99} \\
\cline{3-7}
& & \cellcolor{blue!12}{\textbf{With FL}} & \cellcolor{blue!12}{\textbf{0.83}} & \cellcolor{blue!12}{\textbf{{1.0}}} & \cellcolor{blue!12}{\textbf{0.91}} & \cellcolor{blue!12}{\textbf{0.99}}  \\
\hline

\multirow{6}{*}{\textbf{Node-level}} & \multirow{2}{*}{\textbf{Cadets-E3}} & \textbf{Without FL} & \textbf{{0.97}} & \textbf{{0.99}} & \textbf{{0.98}} & \textbf{{0.99}} \\
\cline{3-7}
& & \cellcolor{blue!12}{\textbf{With FL}}& \cellcolor{blue!12}{\textbf{0.95}} & \cellcolor{blue!12}{\textbf{0.99}} & \cellcolor{blue!12}{\textbf{0.97}} & \cellcolor{blue!12}{\textbf{0.99}} \\
\cline{2-7}
& \multirow{2}{*}{\textbf{Theia-E3}} & \textbf{Without FL }& \textbf{0.98} & \textbf{0.99} & \textbf{0.99} & \textbf{0.99} \\
\cline{3-7}
& & \cellcolor{blue!12}{\textbf{With FL}} & \cellcolor{blue!12}{\textbf{0.97}} & \cellcolor{blue!12}{\textbf{0.99}} & \cellcolor{blue!12}{\textbf{0.99}} & \cellcolor{blue!12}{\textbf{0.99}}\\
\cline{2-7}
& \multirow{2}{*}{\textbf{Trace-E3}} & Without FL & {0.99} & {0.99} & {0.99} & {0.99} \\
\cline{3-7}
& & \cellcolor{blue!12}{\textbf{With FL}} & \cellcolor{blue!12}{\textbf{0.99}} & \cellcolor{blue!12}{\textbf{0.99}} & \cellcolor{blue!12}{\textbf{0.99}} & \cellcolor{blue!12}{\textbf{0.99}}  \\
\hline

\end{tabular}
\end{table}

We see that our model scores high-performance across different datasets, but applying FL reduces the precision in some cases, as in our simulation strategy we are splitting the same dataset over multiple clients, which reduces its quality.

We also compare the average execution time of our solution before and after applying FL across two environments: CPU-based and GPU-based. The results are detailed in Table \ref{tab:time-optimisation}.

\begin{table}[h]
\centering
\caption{Comparison of execution time per execution environment}
\label{tab:time-optimisation}
\vspace{2mm}

\begin{tabular}{c c c c c}%
\cline{2-5}
& \multicolumn{2}{c}{\textbf{Execution time without FL}} &  \multicolumn{2}{c}{\textbf{Execution time with FL}} \\
\hline

\multirow{2}{*}{\textbf{Dataset}} & \multirow{2}{*}{\textbf{On CPU (s)}} & \multirow{2}{*}{\textbf{On GPU (s)}} & \multirow{2}{*}{\textbf{On CPU (s)}} & \multirow{2}{*}{\textbf{On GPU (s)}} \cr \\
\hline
{\textbf{Cadets-E3}} & {1510.2} & {151.2} & \cellcolor{blue!12}{\textbf{322}} & \textbf{67}  \\
\hline 
\textbf{Cadets-Unic} & {25} & {12} & \cellcolor{blue!12}{\textbf{9}} & {\textbf{8}}   \\
\hline
\textbf{Clearscope-E3} & {14} & {6} & \cellcolor{blue!12}{\textbf{5}} & {\textbf{3}}   \\
\hline
\textbf{SC-2} & {4489.2} & {445.8} & \cellcolor{blue!12}{\textbf{427}} & {\textbf{218}}   \\
\hline
\textbf{Streamspot} & {865.8} & {210} & \cellcolor{blue!12}{\textbf{278}} & {\textbf{93}}   \\
\hline
\textbf{Theia-E3} & {852} & {55} & \cellcolor{blue!12}{\textbf{285}} & {\textbf{24}}   \\
\hline
\textbf{Trace-E3} & {3028.8} & {307.2} & \cellcolor{blue!12}{\textbf{465}} & {\textbf{93}}   \\
\hline
\textbf{Wget} & {369} & {34} & \cellcolor{blue!12}{\textbf{67}} & {\textbf{14}}  \\
\hline
\textbf{Wget-HourLong} & {129} & {60} & \cellcolor{blue!12}{\textbf{62}} & {\textbf{24}}   \\
\hline
\end{tabular}
\end{table}

From the data, it is evident that running the solution on a GPU significantly accelerates the execution time, achieving up to a \textbf{10-fold} speed increase compared to execution on a CPU. This highlights the crucial role of parallelization in enhancing the performance of the working environment.
Moreover, the integration of Federated Learning further optimizes execution time. Even when using a CPU, FL reduces the execution time by \textbf{five} to \textbf{ten times} compared to the model without FL, bringing the performance close to that of GPU-based execution. Although a slight reduction in precision was observed in certain datasets, such as SC-2 \citep{han2020unicorn} and Cadets-E3 \citep{githubGitHubDarpai2oTransparentComputing}, after applying FL, the trade-off resulted in a substantial reduction in execution time. For example, the average training and detection time decreased from 1254.78 seconds ($\approx$ 21 minutes) to 213 seconds ($\approx$ 4 minutes), representing an at least \textbf{5-fold} improvement in efficiency.

We compare the performance of our solution—both with and without FL—against state-of-the-art GNN-based IDS in detecting Advanced Persistent Threats. The comparisons for Graph-level datasets are presented in Table \ref{tab:graphlevel-comp}, while those for Node-level datasets are shown in Table \ref{tab:nodelevel-comp}.

\begin{table}[h]
\centering
\caption{Comparison with state-of-the-art on Graph-level datasets}
\label{tab:graphlevel-comp}
\vspace{2mm}

\begin{tabular}{c c c c c c c}%
\hline

\multirow{2}{*}{\textbf{Dataset}} & \multirow{2}{*}{\textbf{IDS}} & \multirow{2}{*}{\textbf{Precision}} & \multirow{2}{*}{\textbf{Recall}} & \multirow{2}{*}{\textbf{F1-Score}} &  \multirow{2}{*}{\textbf{AUC}} & \multirow{2}{*}{\textbf{FP\%}} \cr \\
\hline

\multirow{3}{*}{\textbf{Clearscope-E3}} & Kairos & 0.714 & 1.0 & 0.83 & 0.991 & 1.68 \\
\cline{2-7}
& \textbf{Ours} & \textbf{1.0} & \textbf{1.0} & \textbf{1.0} & \textbf{1.0} & \textbf{0}\\
\cline{2-7}
& \cellcolor{blue!12}\textbf{Ours-FL} & \cellcolor{blue!12}\textbf{1.0} & \cellcolor{blue!12}\textbf{1.0} & \cellcolor{blue!12}\textbf{1.0} & \cellcolor{blue!12}\textbf{1.0} & \cellcolor{blue!12}\textbf{0} \\
\hline

\multirow{3}{*}{\textbf{SC-2}} & Threatrace & 0.923 & 0.960 & 0.941 & - & 4\\
\cline{2-7}
& \textbf{Ours} & \textbf{0.85} & \textbf{0.92} & \textbf{0.88} & \textbf{0.92} & \textbf{8} \\
\cline{2-7}
& \cellcolor{blue!12}\textbf{Ours-FL} & \cellcolor{blue!12}\textbf{0.77} & \cellcolor{blue!12}\textbf{0.96} & \cellcolor{blue!12}\textbf{0.86} & \cellcolor{blue!12}\textbf{0.88} & \cellcolor{blue!12}\textbf{14} \\
\hline

\multirow{6}{*}{\textbf{Streamspot}} & GCA & 1.0 & 0.925 & 0.961 & - & 0 \\
\cline{2-7}
& Magic & 1.0 & 1.0 & 1.0 & 1.0 & 0\\
\cline{2-7}
& Kairos & 1.0 & 1.0 & 1.0 & 1.0 & 0\\
\cline{2-7}
& Threatrace & 1.0 & 1.0 & 1.0 & 1.0 & 0\\
\cline{2-7}
& \textbf{Ours} & \textbf{1.0} & \textbf{1.0} & \textbf{1.0} & \textbf{1.0} & \textbf{0} \\
\cline{2-7}
& \cellcolor{blue!12}\textbf{Ours-FL} & \cellcolor{blue!12}\textbf{1.0} & \cellcolor{blue!12}\textbf{1.0} & \cellcolor{blue!12}\textbf{1.0} & \cellcolor{blue!12}\textbf{1.0} & \cellcolor{blue!12}\textbf{0}\\
\hline

\multirow{4}{*}{\textbf{Wget}} & Magic & 0.96 & 0.96 & 0.96 & 0.962 & 2\\
\cline{2-7}
& Threatrace & 0.926 & 1.0 & 0.98 & - & 4 \\
\cline{2-7}
& \textbf{Ours} & \textbf{1.0} & \textbf{1.0} & \textbf{1.0} & \textbf{1.0} & \textbf{0} \\
\cline{2-7}
& \cellcolor{blue!12}\textbf{Ours-FL} & \cellcolor{blue!12}\textbf{1.0} & \cellcolor{blue!12}\textbf{0.96} & \cellcolor{blue!12}\textbf{0.98} & \cellcolor{blue!12}\textbf{0.97} & \cellcolor{blue!12}\textbf{0}\\
\hline
\end{tabular}
\end{table}

Our solution demonstrates superior performance in Graph-level detection across almost all datasets compared to state-of-the-art IDS. Notably, our model achieved perfect detection of APT attacks in the Wget \citep{DVN/IA8UOS_2018}, Streamspot \citep{manzoor2016fast}, and Clearscope-E3 \citep{githubGitHubDarpai2oTransparentComputing} datasets. However, a slight reduction in performance was observed for the SC-2 dataset \citep{han2020unicorn}, primarily due to the lack of adequate documentation, which complicated the pre-processing phase.

\begin{table}[h]
\centering
\caption{Comparison with state-of-the-art on Node-level datasets}
\label{tab:nodelevel-comp}
\vspace{2mm}

\begin{tabular}{c c c c c c c}%
\hline

\multirow{2}{*}{\textbf{Dataset}} & \multirow{2}{*}{\textbf{IDS}} & \multirow{2}{*}{\textbf{Precision}} & \multirow{2}{*}{\textbf{Recall}} & \multirow{2}{*}{\textbf{F1-Score}} &  \multirow{2}{*}{\textbf{AUC}} & \multirow{2}{*}{\textbf{FP\%}} \cr \\
\hline

\multirow{6}{*}{\textbf{Cadets-E3}} & GHunter & 0.96 & 0.95 & 0.95 & - & -\\
\cline{2-7}
& Magic & 0.94 & 0.99 & 0.97 & 0.99 & 0.22\\
\cline{2-7}
& Threatrace & 0.90 & 0.99 & 0.95 & - & 0.2 \\
\cline{2-7}
& {XFedGraph} & 0.93 & 1.0 & 0.96 & - & - \\
\cline{2-7}
& \textbf{Ours} & \textbf{0.97} & \textbf{0.99} & \textbf{0.98} & \textbf{0.99} & \textbf{0.1} \\
\cline{2-7}
& \cellcolor{blue!12}\textbf{Ours-FL} & \cellcolor{blue!12}\textbf{0.95} & \cellcolor{blue!12}\textbf{0.99} & \cellcolor{blue!12}\textbf{0.97} & \cellcolor{blue!12}\textbf{0.99} & \cellcolor{blue!12}\textbf{0.18}\\
\hline

\multirow{4}{*}{\textbf{Theia-E3}} & GHunter & 0.98 & 0.97 & 0.97 & - & -\\
\cline{2-7}
& Magic & 0.98 & 0.99 & 0.99 & 0.99 & 0.14\\
\cline{2-7}
& Threatrace & 0.87 & 0.99 & 0.93 & - & 0.1 \\
\cline{2-7}
& \textbf{{Ours}} & \textbf{0.98} & \textbf{0.99} & \textbf{0.99} & \textbf{0.99} & \textbf{0.14} \\
\cline{2-7}
& \cellcolor{blue!12}\textbf{Ours-FL} &\cellcolor{blue!12}\textbf{0.97} & \cellcolor{blue!12}\textbf{0.99} & \cellcolor{blue!12}\textbf{0.99} &  \cellcolor{blue!12}\textbf{0.99} & \cellcolor{blue!12}\textbf{0.17} \\
\hline

\multirow{3}{*}{\textbf{Trace-E3}} & Magic & 0.991 & 0.998 & 0.996 & 0.999 & 0.09 \\
\cline{2-7}
& Threatrace & 0.72 & 0.99 & 0.83 & - & 1.1 \\
\cline{2-7}
& \textbf{Ours} & \textbf{0.998} & \textbf{0.999} & \textbf{0.999} & \textbf{0.999} & \textbf{0.01} \\
\cline{2-7}
& \cellcolor{blue!12}\textbf{Ours-FL} & \cellcolor{blue!12}\textbf{0.998} & \cellcolor{blue!12}\textbf{0.999} & \cellcolor{blue!12}\textbf{0.999} & \cellcolor{blue!12}\textbf{0.999} & \cellcolor{blue!12}\textbf{0.01} \\
\hline
\end{tabular}
\end{table}

In Node-level detection, our model performed best on the Theia and Trace datasets from DARPA E3 \citep{githubGitHubDarpai2oTransparentComputing}, underscoring its efficiency in detecting APT attacks within Node-level benchmarks. However, for the Cadets-E3 dataset \citep{githubGitHubDarpai2oTransparentComputing}, a slight degradation in performance was observed when using Federated Learning. While the non-FL version of our model outperformed state-of-the-art systems, the division of data across FL clients resulted in lower precision. In real-world applications, where clients train on their own datasets, this issue would be mitigated.

Overall, the evaluation of Continuum showcased its superior performance in detecting APT attacks, optimizing resources, and surpassing the capabilities of existing models. This demonstrates the potential of our solution in significantly enhancing Intrusion Detection Systems. Further tests and parameter tuning can be performed using our solution, which is available on GitHub\footnote{\url{https://github.com/kamelferrahi/Continuum_FL}}.




\section{Conclusion}
\label{sec:conclusion}
A key contribution of this work is the development of the Spatial-Temporal GNN autoencoder that leverages both spatial and temporal data-modeling. This enables the model to extract deep correlations within the provenance graphs, providing a comprehensive understanding of system behavior. Additionally, we implemented the IDS in a federated-learning environment with homomorphic encryption, ensuring data privacy and security during collaborative training across multiple devices and networks. This approach protects the model’s weights from potential interception and reverse-engineering attacks.

Our experimental results demonstrate the effectiveness of the proposed IDS in accurately detecting APTs while reducing false-positives and resource consumption. By addressing privacy, security, and scalability challenges, our system offers a robust solution for APT detection across different networks and environments. This research showcases the potential of GNNs and federated learning in enhancing IDS capabilities and improving the security of modern computing-environments.

This work also paves the way for several future directions: First, reducing the 
computational and storage overheads associated with generating dynamic-snapshots and processing large provenance-graphs 
especially in real-time or large-scale deployment scenarios is an interesting issue. 
Second, the centralized nature of our FL deployment exposes it to risks such as Single-Point-Of-Failure (SPOF) \citep{techtargetWhatSingle}. These vulnerabilities suggest the need for more advanced, decentralized FL approaches that can mitigate these risks by distributing the aggregation process across multiple nodes, instead of relying on a single server. Lastly, it is also of outmost importance to work on 
dataset availability and quality. In fact, current datasets for APT detection are often outdated, insufficiently documented, and represent only a few stages of APT attacks. 
Moreover, these datasets are not always formatted as graphs and snapshots, necessitating extensive pre-processing and conversion before they can be used for GNN-based Intrusion Detection Systems.

\section*{Acknowledgment}
This work is supported by the French National Research Agency (ANR) under grant ANR-20-CE39-0008 and by Bourg-en-Bresse city.


\newpage
\bibliographystyle{elsarticle-harv}
\bibliography{references.bib}

\end{document}